\documentclass[]{pasj02} 
\usepackage[switch,mathlines]{lineno} 
\usepackage{natbib} 

\jyear{2024}
\Received{}
\Accepted{}

\begin{document} 

\title{PEGASUS (Pfs Emission-line GAlaxy SUrvey with Subaru): I. Three-dimensional large-scale structures at $\bf z > 1$ probed with emission-line galaxies}

\author{
Mariko \textsc{Kubo},\altaffilmark{1,2}\altemailmark\orcid{0000-0002-7598-5292} \email{markubo@kwansei.ac.jp} 
Tadayuki \textsc{Kodama},\altaffilmark{2}\orcid{0000-0002-2993-1576}
Rhythm \textsc{Shimakawa},\altaffilmark{3,4}\orcid{0000-0003-4442-2750} 
Yusei \textsc{Koyama}\altaffilmark{5,6}\orcid{0000-0002-0322-6131} 
Masato \textsc{Onodera}\altaffilmark{5,6,7}\orcid{0000-0003-3228-7264} 
Akio K.  \textsc{Inoue}\altaffilmark{8,9}\orcid{0000-0002-7779-8677} 
Hisakazu \textsc{Uchiyama}\altaffilmark{10}\orcid{0000-0002-0673-0632} 
Tohru \textsc{Nagao}\altaffilmark{11,12}\orcid{0000-0002-7402-5441} 
Ken-ichi \textsc{Tadaki}\altaffilmark{13}\orcid{0000-0001-9728-8909} 
Ronaldo \textsc{Laishram}\altaffilmark{5}\orcid{0000-0002-0322-6131} 
Ryoma \textsc{Wagatsuma}\altaffilmark{2}
Chen \textsc{Nuo}\altaffilmark{2}\orcid{0000-0002-0486-5242} 
Haruka \textsc{Kusakabe}\altaffilmark{14}\orcid{0000-0002-3801-434X} 
Satoshi \textsc{Kikuta}\altaffilmark{15}\orcid{0000-0003-3214-9128} 
Yuma \textsc{Sugahara}\altaffilmark{16,8,9}\orcid{0000-0001-6958-7856} 
Kentaro \textsc{Motohara}\altaffilmark{17,5}\orcid{0000-0002-0724-9146} 
Kazuki \textsc{Daikuhara}\altaffilmark{18}\orcid{0000-0002-2993-1576} 
Ken \textsc{Osato}\altaffilmark{19,20,21,22}\orcid{0000-0002-7934-2569,} 
 and 
Yongming \textsc{Liang}\altaffilmark{23}\orcid{0000-0002-2725-302X} 
}
\altaffiltext{1}{Department of Physics and Astronomy, School of Science, Kwansei Gakuin University, 1 Gakuen Uegahara, Sanda, Hyogo 669-1330, Japan}
\altaffiltext{2}{Astronomical Institute, Tohoku University, 6-3 Aramaki, Aoba-ku, Sendai, Miyagi 980-8578, Japan}
\altaffiltext{3}{Waseda Institute for Advanced Study (WIAS), Waseda University, 1-21-1, Nishi-Waseda, Shinjuku, Tokyo 169-0051, Japan}
\altaffiltext{4}{Center for Data Science, Waseda University, 1-6-1, Nishi-Waseda, Shinjuku, Tokyo 169-0051, Japan}
\altaffiltext{5}{National Astronomical Observatory of Japan, 2-21-1 Osawa, Mitaka, Tokyo 181-8588, Japan}
\altaffiltext{6}{Department of Astronomical Science, The Graduate University for Advanced Studies, 2-21-1 Osawa, Mitaka, Tokyo 181-8588, Japan}
\altaffiltext{7}{Subaru Telescope, National Astronomical Observatory of Japan, National Institutes of Natural Sciences (NINS), 650 North A’ohoku Place, Hilo, HI 96720, USA}
\altaffiltext{8}{Department of Physics, School of Advanced Science and Engineering, Faculty of Science and Engineering, Waseda University, 3-4-1, Okubo, Shinjuku, Tokyo 169-8555}
\altaffiltext{9}{Waseda Research Institute for Science and Engineering, Faculty of Science}
\altaffiltext{10}{Department of Advanced Sciences, Faculty of Science and Engineering, Hosei University, Koganei, Tokyo 184-8584, Japan}
\altaffiltext{11}{Research Center for Space and Cosmic Evolution, Ehime University, 2-5 Bunkyo-cho, Matsuyama, Ehime 790-8577, Japan}
\altaffiltext{12}{Amanogawa Galaxy Astronomy Research Center, Kagoshima University, 1-21-35 Korimoto, Kagoshima 890-0065, Japan}
\altaffiltext{13}{Faculty of Engineering, Hokkai-Gakuen University, Toyohira-ku, Sapporo 062-8605, Japan}
\altaffiltext{14}{Department of General Systems Studies, Graduate School of Arts and Sciences, The University of Tokyo, 3-8-1 Komaba, Meguro-ku, Tokyo, 153-8902, Japan}
\altaffiltext{15}{Nara Prefectural University, 10, Funabashi-cho, Nara, Nara 630-8258}
\altaffiltext{16}{Department of Natural Sciences, Faculty of Science and Engineering, Tokyo City University, 1-28-1 Tamazutsumi, Setagaya, Tokyo 158–8557, Japan}
\altaffiltext{17}{Institute of Astronomy, the University of Tokyo, 2-21-1 Osawa, Mitaka, Tokyo 181-0015, Japan}
\altaffiltext{18}{Institute of Space and Astronautical Science, Japan Aerospace Exploration Agency, 3-1-1, Yoshinodai, Chuou-ku, Sagamihara, Kanagawa 252-5210, Japan}
\altaffiltext{19}{Center for Frontier Science, Chiba University, Chiba 263-8522, Japan}
\altaffiltext{20}{Department of Physics, Graduate School of Science, Chiba University, Chiba 263-8522, Japan}
\altaffiltext{21}{RIKEN Center for Advanced Intelligence Project, Tokyo 103-0027, Japan}
\altaffiltext{22}{Kavli Institute for the Physics and Mathematics of the Universe, The University of Tokyo Institutes for Advanced Study, Chiba, 277-8583, Japan}
\altaffiltext{23}{Institute for Cosmic Ray Research, The University of Tokyo, 5-1-5 Kashiwanoha, Kashiwa, Chiba 277-8582, Japan}




\maketitle

\begin{abstract}
To comprehensively understand the environmental effects on galaxy evolution, it is crucial to precisely map the distribution and thoroughly examine the properties of galaxies, including low-mass building blocks.
Deep narrow-band imaging observations conducted with Hyper Suprime-Cam (HSC) on the Subaru Telescope 
systematically probed the distribution of low-mass star-forming galaxies as 
H$\alpha$, [O {\footnotesize II}], and [O {\footnotesize III}] emission-line galaxies at $0.4\leq z\leq1.6$.
We perform a comprehensive spectroscopic follow-up observation of these emission-line galaxies with narrow-band-based emission-line fluxes $\geq 2\times10^{-17}$ erg s$^{-1}$ cm$^{-2}$ within $\approx1.8$ deg$^2$ area centered on the COSMOS field using `\={O}nohi`ula - Prime Focus Spectrograph (PFS) on the Subaru Telescope.
In this paper, we outline the data product and present the large-scale structures at $z>1$. 
9079 out of 9917 of the targets are observed, and the redshifts of 92\%~of the observed targets are successfully confirmed. 
We survey groups, filamentary large-scale structures, and voids at $z=1.19$, $1.47$, and $1.60$ based on the three-dimensional distribution of $520-920$ [O {\footnotesize II}] emission-line galaxies with stellar masses $M_\star\geq10^{9.5}~M_{\odot}$ at each redshift.
Groups with halo masses $1\times10^{12}-6\times10^{13}~M_{\odot}$ are identified along the filaments, whereas the most massive groups are located at the densest intersections of filaments as predicted by cosmological simulations.
Voids hidden on the surface density maps are robustly extracted on the volume density map.
Our spectroscopic sample provides a powerful basis for quantifying the evolution of galaxies along the diverse environments back in time toward cosmic noon epoch in forthcoming papers.
\end{abstract}


\section{Introduction}

The environmental dependence of galaxy properties has long been known as the morphology--density relation in the local Universe, whereby elliptical and S0 galaxies are predominantly found in the cores of rich clusters, while the general field is dominated by spiral galaxies \citep{1980ApJ...236..351D}, and further presented in colors, star-formation rates (SFR), and star-formation histories (SFH) (e.g., \citealt{1998ApJ...499..589H,1999ApJ...527...54B,2001MNRAS.321...18K,2003ApJ...584..210G, 2004ApJ...601L..29H}). 
Various environmental effects have been proposed to account for the observed environmental dependence of galaxies, such as harassment \citep{1996Natur.379..613M,1998ApJ...495..139M}, thermal evaporation \citep{1977ApJ...211..135C}, ram pressure stripping \citep{1972ApJ...176....1G,2022A&ARv..30....3B}, and starvation or strangulation \citep{1980ApJ...237..692L,2002ApJ...577..651B}, which are closely related to the dynamical evolution of structures involving hot intra-cluster medium (ICM). 
Consequently, the definition of environments and investigation of their impacts on galaxies have been fundamental issues in understanding galaxy evolution.
Beyond the cluster/field definition of environments, galaxies live in a diverse environment consisting of a web-like network of clusters (nodes), filaments, sheets, and voids, called the cosmic web. Numerical simulations have shown that the cosmic web emerged from almost uniform and isotropic small density fluctuations in the early Universe, which have evolved non-linearly via gravitational instability (e.g., \citealt{1996Natur.380..603B}). 
The matter flows along filaments further promote the hierarchical growth of halos at the nodes of the cosmic web, where clusters of galaxies develop.

Along with these evolving environments, galaxies are `pre-processed' in subgroups before they accrete onto larger groups or clusters (e.g., \citealt{1996ApJ...466..104Z,2004PASJ...56...29F,2009MNRAS.400..937M,2019MNRAS.485.2287B}). 
Cosmological numerical simulations have demonstrated that a large fraction of galaxies have experienced pre-processing, which can play critical roles in galaxy evolution (e.g., \citealt{2012MNRAS.423.1277D,2013MNRAS.432..336W,2015MNRAS.447..969B, 2021MNRAS.500.4004D}).
Galaxies can be pre-processed in filaments where the shock-heated gas along filaments \citep{2006ApJ...650..560C} can induce ram-pressure stripping (e.g., \citealt{2017MNRAS.466.4692K,2018ApJ...852..142C,2024MNRAS.528.4139H,2025ApJ...993L..14L}). 
Continuous cold gas flows along filaments can feed galaxies \citep{2014ApJ...796...51D}, and the compression of gas in galaxies passing through filaments can enhance star formation in galaxies \citep{2018MNRAS.480.3152V,2019MNRAS.487.2278V}. 
The dependence of mass and star-formation activities (e.g., \citealt{2017MNRAS.465.3817M,2018MNRAS.474..547K,2021MNRAS.505.4920W}) and morphologies (e.g., \citealt{2013ApJ...779..160Z,2015MNRAS.450.2195S,2026A&A...711A..36E}) of galaxies on the location or distance from filaments, and the alignments of spins aligned with filaments were reported (e.g., \citealt{2013ApJ...775L..42T,2013MNRAS.428.1827T,2025PASJ...77..389T}).
Cosmological numerical simulations have predicted that such structural alignments of galaxies are imprints of the assembly histories of galaxies along the cosmic web 
(e.g., \citealt{2012MNRAS.427.3320C,2018MNRAS.481.4753C,2014MNRAS.444.1453D,2020MNRAS.493..362K}).

Investigating the environmental effects along filaments in detail is, however, challenging because the number density of galaxies along filaments is lower than that of clusters/groups. 
To probe the distribution of filaments, we need to identify the distribution of a vast number of galaxies, both massive and popular low-mass ones, with wide-field observations.
Furthermore, the environmental dependence must be examined alongside the mass dependence, as both mass and environment control galaxy evolution. 
In the local Universe, the observed characteristics of galaxies are well separated into mass and environmental dependencies, with luminous/massive galaxies preferentially observed in overdense regions. At a fixed luminosity/mass, a weaker environmental dependence remains 
(e.g., \citealt{2003ApJ...592..819B,2004ApJ...615L.101B,2004MNRAS.353..713K,2010ApJ...721..193P}).
Internal processes, such as active galactic nuclei (AGN) feedback, are thought to play an important role in quenching star formation in massive galaxies (e.g., \citealt{2005Natur.433..604D,2008ApJS..175..356H}), whereas stellar feedback has a strong impact on low-mass galaxies.
For halos with masses exceeding $10^{12}~M_\odot$, stabilization of the disks results in quenching of their star formation activities (morphological quenching; \citealt{2009ApJ...707..250M}).
As external processes, environmental effects have a large impact on low-mass halos with shallow potential wells.

\begin{longtable}{lcccccccc}
  \caption{Summary of the O2Es, O3Es and HAEs }\label{tab:description}
\hline\noalign{\vskip3pt} 
  Band & Depth\footnotemark[$*$] & Redshift & $\Delta$d \footnotemark[$\dag$] & Line &  $N_{\rm targets}$ & $N_{\rm observed}$ & $N_{\rm \geq 1 line}$\footnotemark[$\ddag$] & $N_{\rm \geq 2 lines}$\footnotemark[$\ddag$] \\   [2pt] 
    & (mag) &  &  (cMpc) &  &  &  &  &  \\   [2pt] 
\hline\noalign{\vskip3pt} 
\endfirsthead      
\hline\noalign{\vskip3pt} 
  Name & Value1 & Value2 & Value3 & Value4 & Value5 & Value6 & Value7 & Value8 \\  [2pt] 
\hline\noalign{\vskip3pt} 
\endhead
\hline\noalign{\vskip3pt} 
\endfoot
\hline\noalign{\vskip3pt} 
\multicolumn{2}{@{}l@{}}{\hbox to0pt{\parbox{160mm}{\footnotesize
\hangindent6pt\noindent
\hbox to6pt{\footnotemark[$*$]\hss}\unskip%
The depth of the narrow-band images. We refer \citet{2020PASJ...72...86H} for {\it NB527/NB718/NB816/NB921/NB973} and \citet{2019PASJ...71..114A} for {\it NB1010}.
\hbox to6pt{\footnotemark[$\dag$]\hss}\unskip%
The line-of-sight coverage of the sample, measured in comoving Mpc, corresponds to the effective bandpass.\\
\hbox to6pt{\footnotemark[$\ddag$]\hss}\unskip%
$N_{\rm \geq 1 line}$ and $N_{\rm \geq 2 line}$ refer to objects identified by the PFS with one emission-line and two or more emission-lines, respectively. The numbers in brackets indicate objects confirmed within the redshift range anticipated from the photometric classification in \citet{2020PASJ...72...86H}. \\
\hbox to6pt{\footnotemark[$\S$]\hss}\unskip%
Further details regarding {\it NB1010} NBEs will be provided in Uchiyama et al. submitted.
}\hss}} 
\endlastfoot 
{\it NB527} & 26.32 & 0.41 & 123 & [O {\footnotesize II}] & 477 & 434 & 394 (389) & 324 (324) \\ 
{\it NB718} & 25.61 & 0.43 & 51 & [O {\footnotesize III}] & 861 & 771 & 762 (758) & 748 (748) \\
            &       & 0.92 & 52 & [O {\footnotesize II}] & 881 & 835 & 825 (813) & 802 (794) \\
{\it NB816} & 25.58 & 0.63 & 47 & [O {\footnotesize III}] & 500 & 448 & 439 (427) & 434 (423) \\
            &       & 1.19 & 45 & [O {\footnotesize II}] & 1002 & 932 & 912 (903) & 749 (743) \\
{\it NB921} & 25.39 & 0.40 & 49 & H$\alpha$ & 510 & 426 & 417 (415) & 410 (410) \\
            &       & 0.84 & 49 & [O {\footnotesize III}] & 2026 & 1950 & 1927 (1897) & 1895 (1876) \\
            &       & 1.47 & 47 & [O {\footnotesize II}] & 1000 & 968 & 946 (920) & 699 (676) \\
{\it NB973} & 24.63 & 0.48 & 48 & H$\alpha$ & 850 & 721 & 693 (687) & 688 (682) \\
            &       & 0.94 & 48 & [O {\footnotesize III}] & 845 & 744 & 709 (690) & 665 (652) \\
            &       & 1.60 & 45 & [O {\footnotesize II}] & 834 & 710 & 608 (520) & 214 (150) \\
{\it NB1010}\footnotemark[$\S$] & 25.1 & ... & ... & ... & 302 & 291 & ... & ... \\
\end{longtable}

Therefore, to gain a comprehensive understanding of environmental impacts, systematic spectroscopic follow-up observations for the targets including a large number of low-mass galaxies across wide fields are crucial.
Our understanding of large-scale structures at high redshifts is now advancing, facilitated by state-of-the-art wide-field spectroscopic surveys with the Euclid \citep{2025A&A...697A...1E}, the Nancy Grace Roman Space Telescope \citep{2015arXiv150303757S}, 
and the new multi-object optical-NIR fiber spectrographs on 10 m class ground-based telescopes, such as the `\={O}nohi`ula Prime Focus Spectrograph (PFS; \citealt{2022SPIE12184E..10T,2024SPIE13096E..05T}) on the Subaru Telescope and the Multi-object Optical and Near-IR Spectrograph (MOONS; \citealt{2020Msngr.180...10C}) on the Very Large Telescope (VLT).
As a precursor, the 360-night PFS Subaru Strategic Program (PFS-SSP) survey started in the S25A semester \citep{2014PASJ...66R...1T}.

However, investigating the environmental effects along cosmic web at $z>1$, the epoch toward cosmic noon, in detail is still challenging within the framework of the above large surveys; it is necessary to probe the distributions of galaxies with stellar masses $M_{\star}<10^{10}~M_\odot$, fainter than the magnitude limit of the target selection of the PFS-SSP survey, with high completeness.
Furthermore, we need to detect relatively faint emission-lines, such as H$\beta$ and [N {\footnotesize II}]$\lambda6550,6585$, which are not always detectable in the framework of the PFS-SSP, to perform emission-line diagnostics robustly.
To complement the PFS-SSP survey, we launched an open-use PFS spectroscopic follow-up program of narrow-band emission-line galaxies (NBEs) selected with Hyper Suprime-Cam (HSC) on the Subaru Telescope. Using the deep and wide narrow-band (NB) images from the HSC Subaru Strategic Program (HSC-SSP; \citealt{2018PASJ...70S...4A}) deep and ultra-deep (UD) layers, and the Cosmic HydrOgen Reionization Unveiled with Subaru (CHORUS; \citealt{2020PASJ...72..101I}) surveys, \citet{2018PASJ...70S..17H,2020PASJ...72...86H} systematically selected H$\alpha$, [O {\footnotesize III}], and [O {\footnotesize II}] emitters at $z\leq 1.6$.
NBEs are representative of typical star-forming galaxies (SFGs) within this redshift range, making them ideal for characterizing large-scale distribution of galaxies (e.g., \citealt{2016ApJS..226....5L, 2018PASJ...70S..21K,2021PASJ...73.1186O,2024ApJ...964L..33L,2025PASJ...77..389T}). 
As the fluxes of H$\alpha$, [O {\footnotesize III}], and [O {\footnotesize II}] are approximately estimated from the narrow-band excesses, it is possible to effectively plan the required exposure times for the targets, including low-mass galaxies. 
Along with our program, named Pfs Emission-line GAlaxy SUrvey with Subaru (PEGASUS), we performed systematic and comprehensive spectroscopic observations concentrated on a sample of NBEs in the COSMOS HSC-SSP UD field during S25A, the first semester of the PFS scientific operation, and in the SXDS HSC-SSP UD field during the S25B semester.

In this first paper of the PEGASUS, we present a summary of the S25A observations and the identification of three-dimensional large-scale structures at $z>1.1$. The structure of this paper is as follows: Section 2 provides a summary of the observations conducted. Section 3 details the data analysis methods used in this study. Section 4 presents the results, Section 5 presents the large-scale structures probed by the PEGASUS, and Section 6 discusses these results. Throughout this paper, we use the cosmological parameters of $H_0 = 70$ km s$^{-1}$ Mpc$^{-1}$, $\Omega_M = 0.3$, and $\Omega_\Lambda = 0.7$. The AB magnitude system \citep{1983ApJ...266..713O} is employed.

\section{Observation} \label{sec:observation}

We conducted a queue-mode observation for the NBEs at $0.4 \leq z \leq 1.6$ in the $\approx1.8$ deg$^2$ survey area centering on the COSMOS UD field using the PFS \citep{2022SPIE12184E..10T,2024SPIE13096E..05T,2026arXiv260614012T} on the Subaru Telescope in the S25A semester (S25A-058QN, PI: M. Kubo).
PFS is the new multi-object fiber spectrograph at the prime focus of the Subaru Telescope, which began scientific operations in the S25A semester.
It operates 2394 science fibers and 96 fiducial fibers simultaneously within a $\sim1.25$ deg$^2$ field of view (FoV).
The blue ($380-650$ nm), red ($630-970$ nm), and near-infrared (NIR) ($940-1260$ nm) arms of the PFS take spectra at $380-1260$ nm simultaneously.
The resolving powers for the blue, red (low-resolution mode), and NIR arms are 
$R\sim2500$ (at 500 nm), $\sim3000$ (at 800 nm), and $\sim4500$ (at 1100 nm), respectively.
The spectral resolutions for the blue, red (low-resolution mode), and NIR arms are $\sim2.1$, 2.7, and 2.4 \AA, respectively, enough to resolve the emission-line profiles of typical star-forming galaxies. 
In this section, we describe the targets and requested queue (section \ref{ssec:targets}) and the observation executed (section \ref{ssec:s25aobs}) in the S25A semester.

\subsection{Targets and requested exposure time} \label{ssec:targets}

Our targets were the NBEs at $0.4\leq z \leq 1.6$ in the COSMOS field selected by \citet{2020PASJ...72...86H}.
They constructed a NBE catalog based on the HSC-SSP \citep{2018PASJ...70S...4A} public data release 2 (PDR2; \citealt{2019PASJ...71..114A}) and Cosmic HydrOgen Reionization Unveiled with Subaru (CHORUS) survey \citep{2020PASJ...72..101I}.
The details of the NBE selection procedure are provided in \citet{2020PASJ...72...86H}. 
In summary, NBEs were selected using {\it NB527}, {\it NB718}, {\it NB816}, {\it NB921}, and {\it NB973}-band images in the HSC-SSP COSMOS UD field. 
The $5\sigma$ detection limits of these narrow-band images are $24.63-26.32$ mag, which can recover an emission-line with $F_{\rm NB} >1-3\times10^{-17}$ erg s$^{-1}$ cm$^{-2}$, 
where $F_{\rm NB}$ is an emission-line flux measured with a narrow-band in \citet{2020PASJ...72...86H}.
Once narrow-band excess sources were selected via broad-band vs. narrow-band colors, they were classified into H$\alpha$, [O {\footnotesize II}]$\lambda\lambda3727,3730$, and [O {\footnotesize III}]$\lambda\lambda4960,5008$ emitters (HAEs, O2Es, and O3Es)
based on broad-band colors and/or photometric redshifts.
The effective bandwidths of these filters are $70-130$ \AA, which corresponds to a few 1000 km s$^{-1}$ or $45-123$ cMpc at $0.4\leq z\leq 1.6$; if a cluster of galaxies is centered on the transmission curve, all member galaxies can be selected as NBEs.

Table \ref{tab:description} summarizes the number of targets in the S25A semester.
We requested to observe all NBEs (10088, reduced to 9917 after removing duplicates) with $F_{\rm NB}>2\times10^{-17}$ erg s$^{-1}$ cm$^{-2}$ within a region of $\sim1.5$-degree diameter centered on the HSC-SSP COSMOS UD field.
This emission-line flux limit corresponds to an SFR $\sim0.1~M_{\odot}~\text{yr}^{-1}$ for HAEs at $z=0.4$ adopting H$\alpha$ luminosity to SFR relation in \citet{1998ARA&A..36..189K} described as SFR$(M_{\odot} \text{yr}^{-1}) = 7.9\times10^{-42} L(H\alpha)$(erg s$^{-1}$). 
This relation, originally described with \citet{1955ApJ...121..161S} IMF, was converted into the formalism of \citet{2003PASP..115..763C} IMF by applying a multiplication factor of 0.64 used in \citet{2014ARA&A..52..415M}. 
We also assumed a marginal dust attenuation correction of $E(B-V)=0.1$ following the attenuation law of \citet{2000ApJ...533..682C}.
Assuming [O {\footnotesize II}]$\lambda\lambda3727,3730$/H$\alpha=1$, this flux limit corresponds to SFR $\sim2~M_{\odot}~\text{yr}^{-1}$ for O2Es at $z=1.47$. 

We requested exposure times that are optimal to achieve our primary goals: confirm the redshifts of NBEs at $0.4\leq z \leq 1.6$ almost completely and obtain emission-line ratios, such as H$\alpha$/H$\beta$, [O {\footnotesize III}]$\lambda5008$/H$\beta$ and, ideally, [N {\footnotesize II}]$\lambda6585$/H$\alpha$ within the PFS spectral coverage.
We aimed to detect H$\beta$ and [N {\footnotesize II}]$\lambda6585$ emission-lines from NBEs at $z<1.5$ and $z<0.9$, respectively, with $>3\sigma$ significance.
For O2Es at $z\approx1.60$, we only aimed to confirm the redshifts by detecting one or more strong emission-lines like [O {\footnotesize II}]$\lambda\lambda3727,3730$ and [O {\footnotesize III}]$\lambda\lambda4960,5008$.
The expected fluxes of these emission-lines were calculated based on $F_{\rm NB}$ and photometric redshifts in \citet{2020PASJ...72...86H} and empirical line ratios.
The optimal exposure time for each target was calculated using the PFS Spectral Simulator\footnote{https://pfs-etc.naoj.hawaii.edu/etc}. 
$0.25-4.0$ h exposure was requested for each target (Table \ref{tab:description1} in appendix).
In addition to the NBEs in \citet{2020PASJ...72...86H}, we put 302 {\it NB1010} NBEs with {\it NB1010} EW in the rest-frame ($EW_0$) $>100$ \AA~in the target list. 
Such high-EW NBEs are candidate young metal-poor AGNs \citep{2007ApJ...668..853K,2008ApJ...687..133I}.
A two-hour exposure was requested for each {\it NB1010} NBE (see details in Uchiyama et al. submitted). 

In the target list submitted to the observatory, we set a reference arm, which is the most important wavelength range, for each target. For {\it NB527}, {\it NB718}/{\it NB816}/{\it NB921}, and {\it NB973}/{\it NB1010} NBEs, we set the blue, red, and NIR arms, respectively, as reference arms.
We requested queue observing time where $>90$ \% of the targets fulfill their dedicated observing time.

\subsection{S25A observation} \label{ssec:s25aobs}

\begin{figure*}
 \begin{center}
  \includegraphics[width=\linewidth]{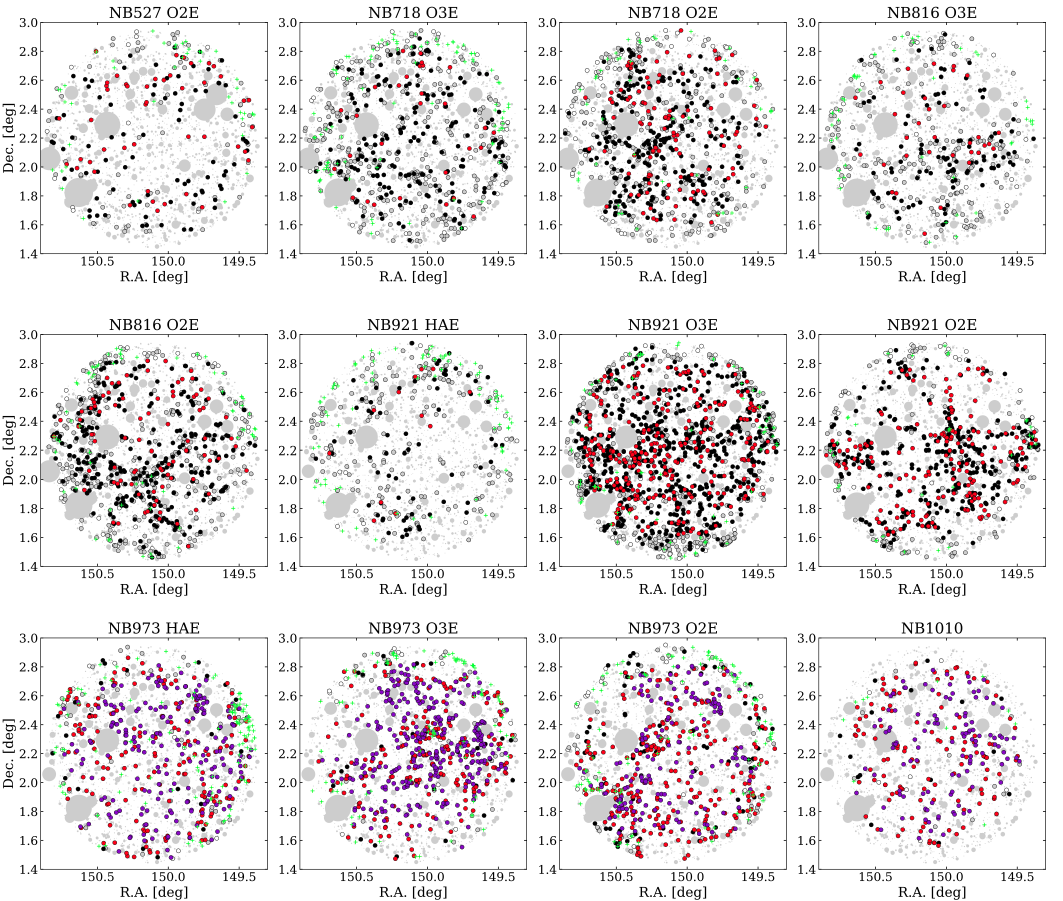} 
 \end{center}
\caption{The sky distribution of the targets. The panels show the sky distributions of {\it NB527} O2Es, {\it NB718} O3Es and O2Es, {\it NB816} O3Es and O2Es, {\it NB921} HAEs, O3Es and O2Es, {\it NB973} HAEs, O3Es and O2Es, and {\it NB1010} NBEs from the top left to the bottom right.
The white, gray, black, red, and violet filled points show the targets that fulfill $<33$, $34-66$, $67-99$, $100-200$, and $>200$ \% of the requested exposure time (not in EET), respectively.
The green crosses show the targets with no visit.
The gray-filled regions show the bright star masks applied in \citet{2020PASJ...72...86H}.
}\label{fig:skydist1}
\end{figure*}

The observations were conducted in queue mode from 26 March to 2 April 2025.
The single exposure unit, called a visit, was 450 s in the S25A semester.
The PFS fibers were shared by multiple programs, and the observatory controlled the visits for each target. 
The queue completion rate of our program was 105.07\%, which was calculated based on the effective exposure time (EET).
EETs were defined by considering the actual observing conditions.
If the actual observing condition of a visit is better than the usual condition, EET for this visit is larger than its nominal exposure time.
The number of observed ($\geq1$ visits) NBEs is listed in Table \ref{tab:description}.
In total, 9230/10088 objects in the target list were observed. 
$84-97$ \% of NBEs were observed, while the fractions of observed NBEs were lower for NBEs at lower redshifts and {\it NB973} NBEs.
Figure \ref{fig:skydist1} presents the completion rate of the requested exposure time. 
Even in the densest regions, most of the targets met more than two-thirds of the requested total integration time.
Based on EETs, most of the targets met the requested observing time. 
The targets with no visits were located mostly at the edge of the target field.

Note that {\it NB973} and {\it NB1010} NBEs had a larger number of visits than those of the other NBEs. 
This may be because the reference arms of {\it NB973} and {\it NB1010} NBEs were set to the NIR arm, which is known to have persistence problems in the S25A semester \citep{2026arXiv260614012T}. 
Duplicates of NBEs, especially {\it NB527} O2Es and {\it NB921} HAEs, were not removed from the submitted target list because their identification was independent in \citet{2020PASJ...72...86H}. 
As observed targets, 147 {\it NB527} O2Es were duplicated with {\it NB921} HAEs.
In addition, there were three {\it NB921} O3Es duplicating with {\it NB527} O2Es and one {\it NB718} O2E duplicating with {\it NB921 O3E} in the observed target list.
Thus, in total, 
151 (171) NBEs among the observed (submitted) target list were duplicated. 
Since we primarily focus on the survey performance of the PEGASUS in this paper, we simply used the spectra taken with longer exposure for the duplicated objects.
In summary, 9079 unique objects were observed in the S25A semester. 

\begin{figure*}
 \begin{center}
  \includegraphics[width=\linewidth]{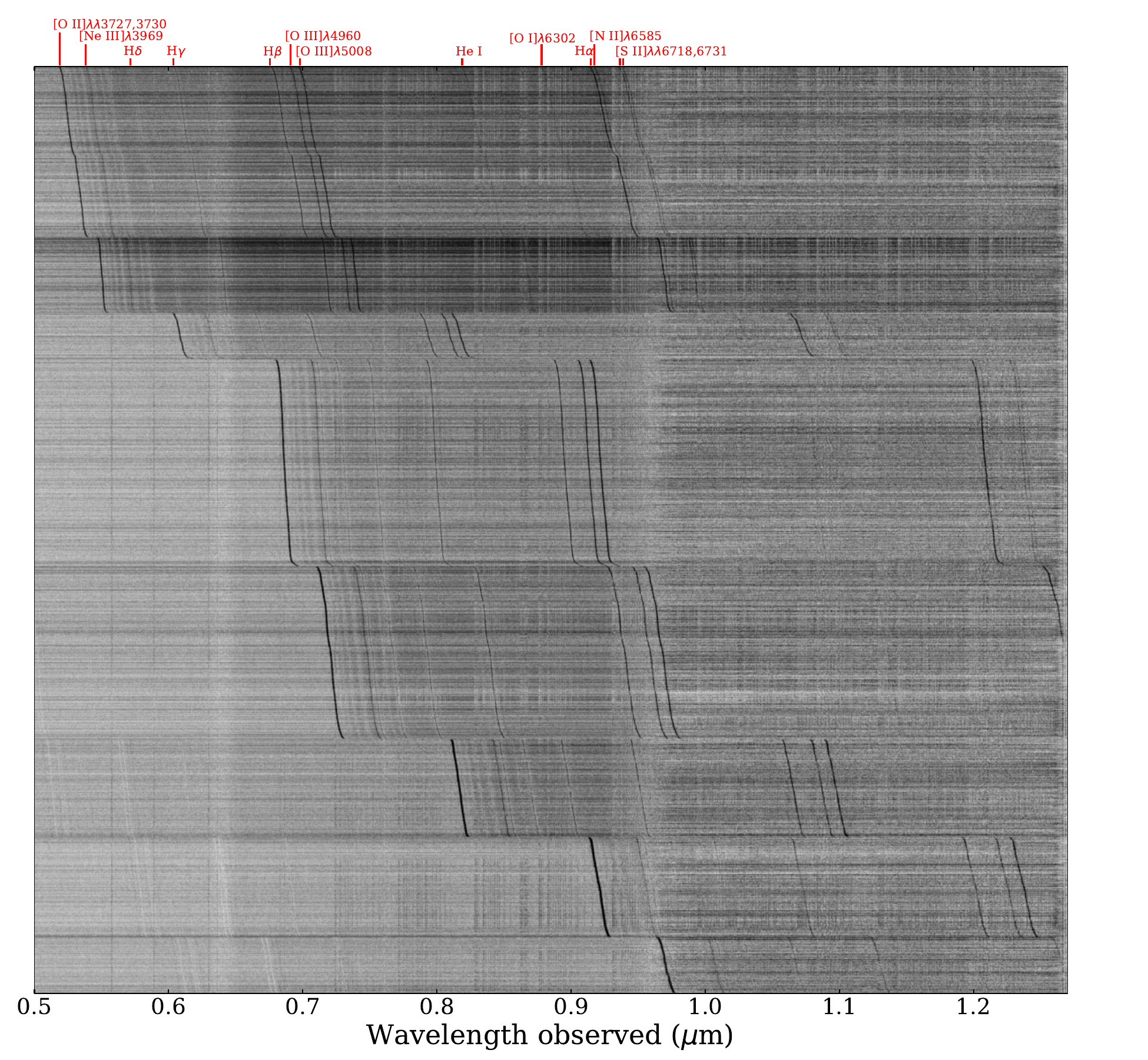} 
 \end{center}
\caption{The PFS spectra of the confirmed targets. The spectra are arranged in the order of redshift. The red colored text notes major emission-lines.
}\label{fig:2dimage}
\end{figure*}

\begin{figure*}
 \begin{center}
  \includegraphics[width=0.8\linewidth]{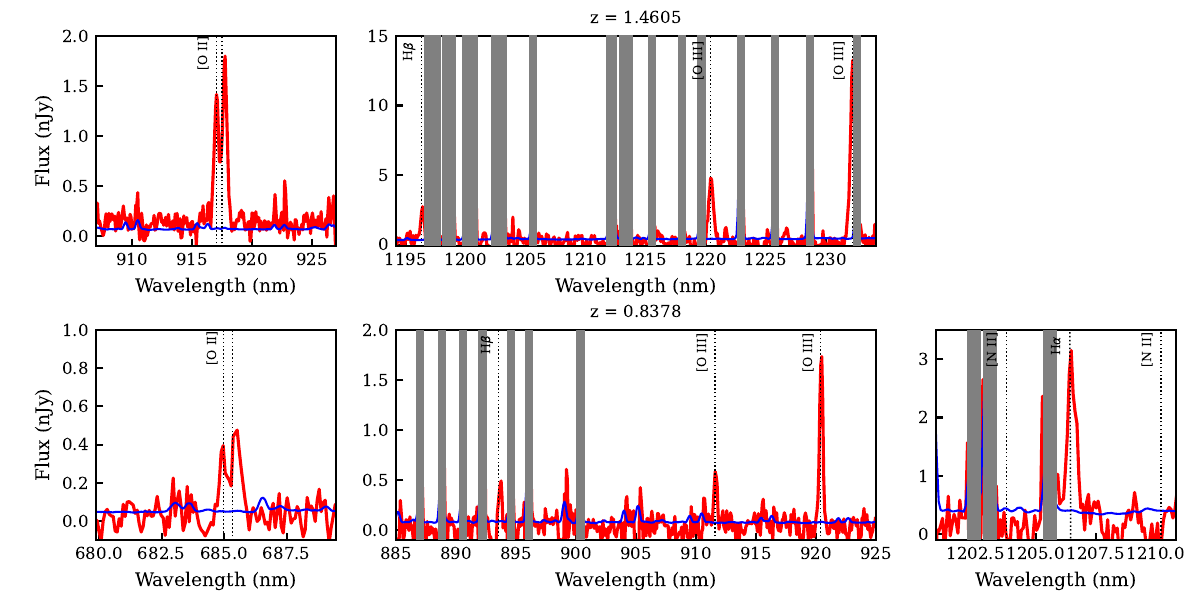} 
 \end{center}
\caption{Spectra of an O2E at $z\approx1.47$ ({\it top}) and an O3E at $z\approx0.84$ ({\it bottom}).
The red and blue curves show the fluxes and flux errors, respectively.
The gray-shaded area masks the OH airglow.
The dotted vertical lines show the locations of emission-lines.
}\label{fig:spectrum1}
\end{figure*}

\section{Data analysis}
\subsection{Pipeline data reduction}

The raw data were reduced using the PFS data reduction pipeline ({\sf PFS DRP}) at the observatory 
and delivered to us via the PFS science platform (PFS-SP)\footnote{https://hscpfs.mtk.nao.ac.jp/}. 
Here, we used the 2D pipeline ({\sf 2D-DRP}) data product at the first data release (run21\_June2025) delivered in June 2025.
The 2D pipeline version was PIPE2D-1703-7f169e3.
The raw data for each visit in each arm were reduced by removing the basic instrumental signature, calibrating the wavelength, subtracting the sky, calibrating fluxes, and merging the three-arm spectra.
The reduced spectra of multiple visits for an object were then coadded. See details in the {\sf PFS DRP} webpage\footnote{https://subaru-pfs.github.io/pfs\_helpdesk\_tutorial}.
Note that this version of the data reduction pipeline had issues with sky subtraction for the NIR arm (see details in \citealt{2026arXiv260614012T}).
Therefore, the results based on the NIR data in this paper are not yet fully guaranteed.

The FITS image files of three-arm-combined 1D spectra, 
including fluxes, flux errors, and masks for bad pixels, were provided as the {\sf 2D-DRP} data product. 
The following data analyses were performed based on the 1D spectra of the {\sf 2D-DRP} product, primarily using {\sf astropy} in {\sf Python}.
The observatory also provides the 1D pipeline ({\sf 1D-DRP}) data product, which provides, e.g., spectroscopic redshifts measured on the {\sf 2D-DRP} data product, but we did not use them.
We note that the spectroscopic redshifts measured by the {\sf 1D-DRP} had a large scatter from our visually checked spectroscopic redshifts.

\subsection{Redshift and flux estimation}

First, we estimated redshifts using {\sf slinefit}\footnote{https://github.com/cschreib/slinefit}, 
which is a simple software tool for deriving spectroscopic redshifts from spectra. 
We used the fluxes, flux errors, and bad-pixel masks of each spectrum in the {\sf 2D-DRP} data product. 
We ran {\sf slinefit} to fit the spectra within a narrow redshift range ($\Delta z = \pm0.05$) around the photometric redshifts of the NBEs in \citet{2020PASJ...72...86H}. 
We then visually checked whether a model spectrum obtained with {\sf slinefit} successfully fitted the observed spectrum. 
Most of the spectra were well-fitted using {\sf slinefit}, but several spectra failed to fit because of the misclassification of the photometric redshifts and no line detections. 
For such sources, we manually set initial values based on visual inspections of the spectra and reran {\sf slinefit}. 

Using the redshifts obtained with {\sf slinefit} as initial guesses, we re-fitted the emission-line profiles with Gaussian profiles 
by maximum likelifood method performed with MCMC using {\sf emcee}\footnote{https://emcee.readthedocs.io/en/stable/} \citep{2013PASP..125..306F}.
Each emission-line was fitted with a single-component Gaussian profile after subtracting the continuum evaluated around each emission-line.
We adopted fixed line ratios of [O {\footnotesize III}]$\lambda5008$/[O {\footnotesize III}]$\lambda4960 = 3$ and [N {\footnotesize II}]$\lambda6585$/[N {\footnotesize II}]$\lambda6550 = 3$.
First, H$\alpha$ ([O {\footnotesize III}]$\lambda\lambda4960,5008$ doublet for $z>0.9$) emission-line of a spectrum was fitted with a Gaussian profile (two Gaussian profiles for [O {\footnotesize III}]$\lambda\lambda4960,5008$) parameterized by redshift, line width, and flux value. 
Then H$\beta$, [O {\footnotesize II}]$\lambda\lambda3727,3730$, [O {\footnotesize III}]$\lambda\lambda4960,5008$, and [N {\footnotesize II}]$\lambda\lambda6550,6585$ emission-lines of a spectrum were fitted with Gaussian profiles parameterized by redshifts and flux values, adopting a line width measured with H$\alpha$ ([O {\footnotesize III}]$\lambda\lambda4960,5008$). 

We measured the S/N ratio for each emission-line by taking the flux-to-flux error ratio at a 200 km s$^{-1}$ interval centered at each line center. 
For H$\alpha$, [O {\footnotesize III}]$\lambda5008$, and [O {\footnotesize II}]$\lambda\lambda3727,3730$ emission-lines, an emission-line with S/N $>5$ was regarded as a detection. 
For other emission-lines, an emission-line with S/N $>3$ was regarded as a detection. 
The median and standard deviation of the central wavelength of the detected emission-lines were adopted to derive the spectroscopic redshift and error of each target.
Because the data reduction pipeline was not yet complete (\citealt{2026arXiv260614012T}), 
the results based on the spectra, especially at $>0.94~\mu$m (NIR arm), may change in the forthcoming papers.
We will make the catalog publicly available along with subsequent papers based on the future version of reduced data.

\section{Results}
\label{ssec:results}

\begin{figure*}
 \begin{center}
  \includegraphics[width=0.95\linewidth]{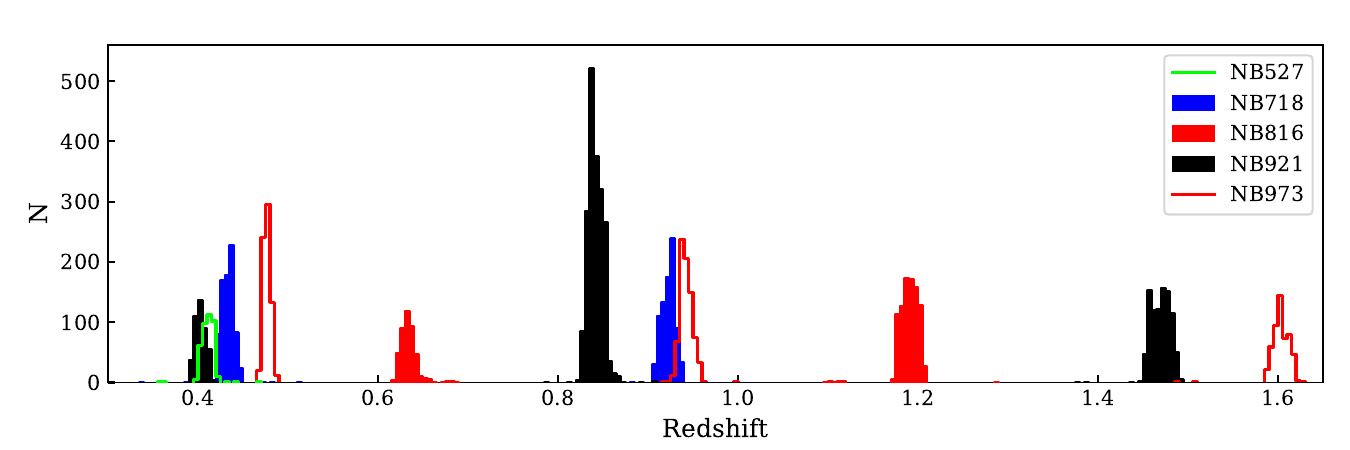} 
 \end{center}
\caption{The redshift distribution of NBEs confirmed by the PEGASUS survey. 
The green solid line, blue-filled, red-filled, black-filled, and red solid line histograms present 
the redshift distributions of the {\it NB527}, {\it NB718}, {\it NB816}, {\it NB921} and {\it NB973} NBEs, respectively.
}\label{fig:zdist0}
\end{figure*}

\begin{figure}
 \begin{center}
  \includegraphics[width=0.95\linewidth]{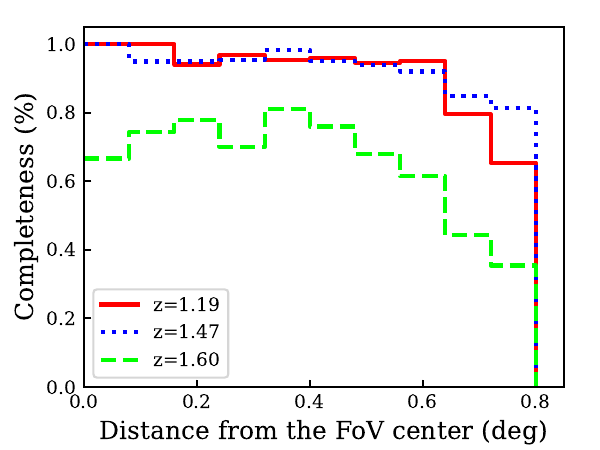} 
 \end{center}
\caption{ The redshift confirmation completeness of O2Es as a function of the distance from the survey area center. 
The red solid, blue dotted, and green dashed histograms show the redshift confirmation completeness for O2Es at $z=1.19$, $ z=1.47$, and $ z=1.60$, respectively.
}\label{fig:completeness_o2es}
\end{figure}

\begin{figure*}
 \begin{center}
  \includegraphics[width=0.95\linewidth]{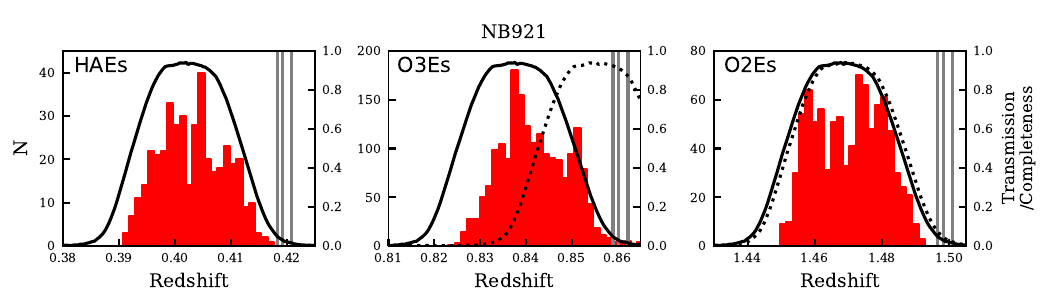} 
 \end{center}
\caption{ The redshift distributions of the {\it NB921} HAEs, O3Es, and O2Es from left to right. 
The red histograms show the redshift distributions of NBEs, excluding outliers. 
The black solid (dotted) curves show the transmission curve of the {\it NB921} filter scaled to H$\alpha$, [O {\footnotesize III}]$\lambda5008$ ($4960$), and [O {\footnotesize II}]$\lambda3730$ ($3727$).
The gray shaded area masks the OH airglows.
}\label{fig:zdist1}
\end{figure*}

\subsection{Redshifts and line detections}
\label{ssec:spectra}

Figure \ref{fig:2dimage} shows the coadd spectra of the confirmed targets arranged in redshift order, and Figure \ref{fig:spectrum1} shows examples of the spectra.
Many of the spectra show Balmer series, [O {\footnotesize II}]$\lambda\lambda3727,3730$, [Ne {\footnotesize III}]$\lambda$3869, [O {\footnotesize III}]$\lambda\lambda4960,5008$ ($z<1.5$), and [S {\footnotesize II}]$\lambda\lambda6718,6733$ ($z<0.85$) emission-lines clearly.
[N {\footnotesize II}]$\lambda\lambda6550, 6585$ were generally weak in each spectrum. 
[O {\footnotesize II}]$\lambda\lambda3727,3730$, which are useful to estimate electron density, were often resolved well.

Figure \ref{fig:zdist0} shows the overall redshift distribution of the successfully confirmed targets, and
Table \ref{tab:description} presents the number of targets successfully confirmed in their redshifts.
Several redshift intervals at $0.4\leq z\leq 1.6$ were sampled with a large number ($N=400-1900$) of NBEs.
The redshifts of $86-99$ \% of the observed targets were successfully confirmed by detecting one or more emission-lines.
The success rate for {\it NB973} NBEs was relatively low because their primary emission-lines were shifted into the NIR arm, where OH airglow and noisy spectra from incomplete sky subtraction prevented line detections.

Figure \ref{fig:completeness_o2es} illustrates the completeness of redshift confirmations for O2Es at $z = 1.19$, $1.47$, and $1.60$ as a function of the distance from the survey area center. 
Completeness was defined as the fraction of O2Es confirmed at these three redshifts among the O2E candidates selected by {\it NB816}, {\it NB921}, and {\it NB973} in the target list, respectively.
O2Es at $z=1.19$ and $1.47$ achieve a completeness of over 90\% throughout the survey area. 
Although the completeness of O2Es at $z=1.60$ is relatively low, their redshift confirmations were performed uniformly within a 0.6-degree radius from the survey area center.
Such a uniform sampling with high completeness is crucial for large-scale structure extractions performed in section \ref{sec:lss}.

$<3$\% of the targets turned out to be emission-line galaxies 
that differed from the photometric classification by \citet{2020PASJ...72...86H}, 
except for {\it NB973} O2Es, which includes $\sim14$\% interlopers.
This demonstrates the robustness of the photometric classification scheme of NBEs.
These NBEs were mostly misclassified as HAEs, O3Es, or O2Es detectable with the same narrow-band image. 
A few outliers were identified as H$\beta$ or [Ne {\footnotesize III}] emitters.
Several targets were not confirmed spectroscopically in this work; the total integration times were too short ($<1$ hour) for two-thirds of them. 
The rest of them met the requested observing time, but all have relatively small narrow-band fluxes $\sim2\times10^{-17}$ erg s$^{-1}$ cm$^{-2}$. Probably, they were false NBEs.

Figure \ref{fig:zdist1} shows the spectroscopic redshift distributions of HAEs, O3Es, and O2Es selected as {\it NB921} NBEs compared to the transmission curve of {\it NB921} filter.
The spectroscopic redshift distributions for the other NBEs are presented in Figure \ref{fig:zdistmerged} in appendix.
The redshift distributions of the HAEs and O2Es align with the centers of the filter transmission curves, while the O3Es exhibit a slight bias toward higher redshifts.
This is because both [O {\footnotesize III}]$\lambda 4960$ and [O {\footnotesize III}]$\lambda 5008$ are relevant for the narrow-band flux excesses.
The OH airglow is encompassed within the redshift coverage of {\it NB973} NBEs but does not significantly disturb the identification of the redshift distribution (Figure \ref{fig:zdistmerged} in appendix).

We checked the duplication of our spectroscopically confirmed galaxies against the COSMOS Spectroscopic Redshift Compilation catalog \citep{2026ApJS..282....6K}.
We selected the objects within $0''.2$ of galaxies in our catalog and with quality flags for the archival data $\geq1$ (50\% confidence) in \citet{2026ApJS..282....6K}.
306 of our spectroscopic sample were confirmed in their redshifts with NIR spectroscopic observations in previous studies.
1813 of our spectroscopic sample were confirmed in their redshifts with optical spectroscopy, where 1246 of them were confirmed by the Dark Energy Spectroscopic Instrument (DESI) survey.
The spectroscopic redshifts were mostly consistent with the compilation catalog by \citet{2026ApJS..282....6K}.
Although this paper does not aim to provide an exhaustive comparison with previous spectroscopic surveys, the spectra taken with other facilities will be useful in correcting fiber flux losses in future work.

\subsection{Flux losses from fibers}
\label{ssec:fluxlimit}

As PFS is a fiber spectrograph, we need to consider the flux losses from the fiber aperture.
Considering the general seeing sizes in the optical at the Subaru Telescope ($\sim 1''$), the apparent sizes of the targets can often exceed the fiber core size of the PFS $\approx1''.1$.
We presented the emission-line fluxes measured with narrow-band ($F_{\rm NB}$) to those obtained using the PFS ($F_{\rm sp}$) in Figure \ref{fig:fluxloss} in appendix.
For HAEs and O3Es, the fluxes of H$\alpha$ and [O {\footnotesize III}]$\lambda5008$ were presented as $F_{\rm sp}$, respectively.
For O2Es, the fluxes of [O {\footnotesize II}]$\lambda\lambda3727,3730$ were presented as $F_{\rm sp}$.
[N {\footnotesize II}]$\lambda\lambda6550, 6585$ emission-lines are generally weak for our targets and do not have a significant impact on the $F_{\rm sp}/F_{\rm NB}$ ratios.
The $F_{\rm sp}/F_{\rm NB}$ ratios for HAEs and O2Es follow the transmission curves.
The $F_{\rm sp}/F_{\rm NB}$ ratios for O3Es exhibited wider redshift distributions corresponding to the transmission curves scaled to [O {\footnotesize III}]$\lambda4960$ and [O {\footnotesize III}]$\lambda5008$.

There is a considerable dispersion of $F_{\rm sp}$/$F_{\rm NB}$ even at the plateau of the transmission curves.
Most of the targets have $F_{\rm sp}$/$F_{\rm NB}<1$, i.e., $F_{\rm NB}$ is larger than $F_{\rm sp}$. 
Thus, the variation in $F_{\rm sp}$/$F_{\rm NB}$ is mainly attributed to the flux losses from the PFS fibers. 
The non-negligible number of O2Es have $F_{\rm sp}$/$F_{\rm NB}>1$. 
It may be due to the large uncertainties in the subtraction of the continuum from narrow-band images, as [O {\footnotesize II}]$\lambda\lambda3727,3730$ is close to the Balmer break.
The $F_{\rm sp}$/$F_{\rm NB}$ of {\it NB973} NBEs have a lot scatter.
This can be attributed to the flux calibration of the NIR arm.

\subsection{Emission-line flux and stellar mass distributions}
\label{ssec:population}

\begin{figure}
 \begin{center}
  \includegraphics[width=0.95\linewidth]{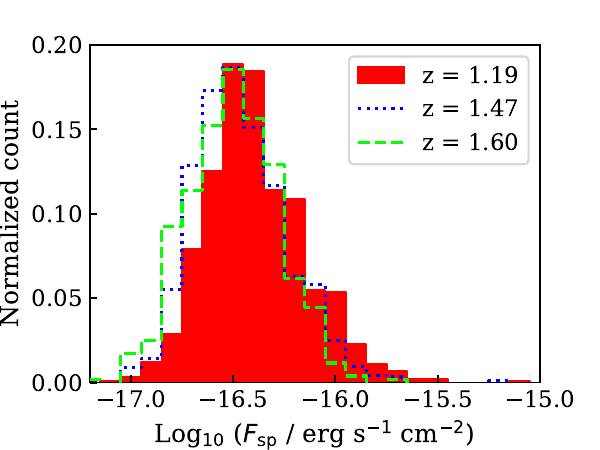} 
 \end{center}
\caption{ [O {\footnotesize II}]$\lambda\lambda3727,3730$ emission-line flux ($F_{\rm sp}$) distributions of O2Es measured with the PFS. 
The red-filled, blue-dotted, and green-dashed histograms present O2Es at $z=1.19, 1.47,~\&,1.60$, respectively, which are detected with $>3\sigma$ significance.
}\label{fig:fluxdist}
\end{figure}

\begin{figure}
 \begin{center}
  \includegraphics[width=0.95\linewidth]{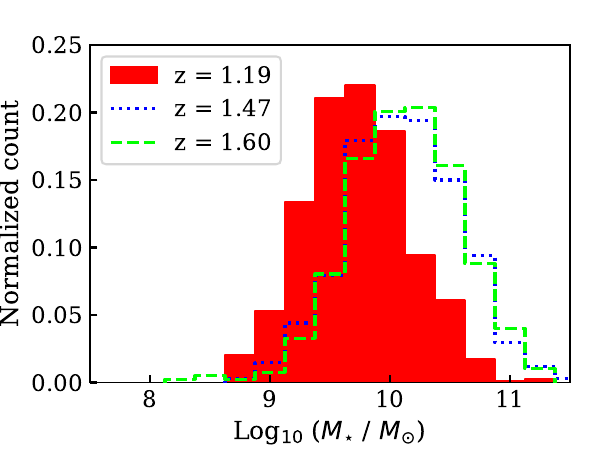} 
 \end{center}
\caption{ Similar to Figure \ref{fig:fluxdist}, but shows the stellar mass distributions of O2Es.
}\label{fig:smdist1}
\end{figure}

Figure \ref{fig:fluxdist} shows the [O {\footnotesize II}]$\lambda\lambda3727,3730$ emission-line flux distributions of O2Es at $z=1.19,~1.47$ and $1.60$.
O2Es with [O {\footnotesize II}]$\lambda\lambda3727,3730$ emission-line fluxes $\gtrsim 2\times 10^{-17}$ erg s$^{-1}$ cm$^{-2}$ were identified.
Figure \ref{fig:smdist1} is similar to Figure \ref{fig:fluxdist} but shows the distribution of stellar masses.
We utilized the stellar masses estimated by \citet{2020PASJ...72...86H}, which were calculated with the redshifts corresponding to the center of each narrow-band filter.
The stellar masses of O2Es range from $10^8$ to $10^{11.5}~M_{\odot}$ and the O2Es with stellar masses $\geq 10^{9.5}~M_{\odot}$ were likely identified with high completeness.

\section{Groups and large-scale structures probed with PEGASUS}
\label{sec:lss}

In this first paper of the PEGASUS, we present the most detailed map of three-dimensional large-scale structures probed with O2Es at $z=1.19$, $1.47$, and $1.60$.
Clusters and filaments have been surveyed in the COSMOS field based on photometric redshifts in previous studies \citep{2018MNRAS.474.5437L,2019MNRAS.489.5695D,2024ApJ...976..154K}.
O2Es, which represent low-mass SFGs with a large number density, serve as excellent probes of large-scale structures, as demonstrated by 
cosmological numerical simulations (e.g., \citealt{2018MNRAS.474.4024G,2020MNRAS.498.1852G,2020MNRAS.497.5432F,2023MNRAS.519.1771O}) and observations (e.g., \citealt{2021PASJ...73.1186O,2025MNRAS.539.2323I,2025PASJ...77..389T,2026arXiv260708453M}). 
Here, the large-scale structures were investigated in the volume number density, and extractions of groups, filaments, and voids, using O2Es.

\subsection{Methods}
\label{sec:method}

\subsubsection{Completeness of O2Es}
\label{sssec:method_mask}

First, we assessed the completeness of the sample selection. 
We observed all NBEs with $F_{\rm NB}>2\times10^{-17}\rm erg~s^{-1}~cm^{-2}$ in the target field.
According to Figure \ref{fig:completeness_o2es}, a radius of $\approx0.6$ deg from the survey area center was uniformly sampled. 
The incompleteness in the redshift direction depends on the narrow-band photometric selection. 
In case of our targets with $F_{\rm NB}>2\times10^{-17}\rm erg~s^{-1}~cm^{-2}$, O2Es at $z=1.179-1.205$ and $z=1.455-1.485$ should be detected with $>10\sigma$ significance on the {\it NB816} and {\it NB921}-band images of the survey area, respectively, i.e., NBEs at this limit were nearly completely selected, considering the depth of the HSC-SSP PDR2 \citep{2019PASJ...71..114A}.
In case of {\it NB973}, O2Es at $z=1.587-1.617$ can be detected with $>5\sigma$ significance.
This expectation agrees with the tendencies of $F_{\rm SP}/F_{\rm NB}$ and redshift distributions presented in Figure \ref{fig:zdist1}, and Figures \ref{fig:zdistmerged} and \ref{fig:fluxloss} in appendix.

The bright stars were masked in the catalog by \citet{2020PASJ...72...86H}. 
We checked the impact of the bright star masks on the structural measurement. 
We calculated the expected number of O2Es in the masked regions by multiplying the masked area by the volume number density of O2Es, incorporating the above target completeness.
The mean volume number density of O2Es was estimated by counting O2Es within a 0.6 deg radius of the survey area center and the above redshift intervals, excluding the area masked by bright stars.
In total, $30-50$ O2Es are expected in the masked area within a 0.6 deg radius of the survey area center of each redshift slice.
Thus, the bright star masks marginally affect the density map.

In summary, O2Es were uniformly sampled within a 0.6 deg radius from the survey area center and in the redshift intervals $z=1.179-1.205$ and $z=1.455-1.485$ 
while the completeness of O2Es at $z=1.587-1.617$ is lower but nearly uniformly sampled.
Considering a typical velocity width of a massive cluster $\pm1000$ km s$^{-1}$,
clusters of galaxies at the redshift interval $z=1.184-1.200$, $z=1.460-1.480$ and $z=1.592-1.612$ can be completely sampled.
We used all the O2Es at each redshift slice in the following analyses.
The variation in completeness may not significantly affect the location of groups and filaments
because their extraction was performed based on local density.
However, the volume number density, the richness of the groups, and the significance of filaments can be underestimated outside the above area.
In addition, O2Es are biased to low-mass star-forming galaxies as presented in Figure \ref{fig:smdist1}.
As we describe below, the cores of massive groups can already be dominated by massive passively evolving galaxies, and O2Es do not always probe the galaxy distribution at such regions well.

\subsubsection{Volume number density estimation}
\label{sssec:method_3d_nd}

The volume number density distributions were defined with the Nth nearest-neighbor method. 
First, the celestial coordinates and redshifts of the O2Es were transformed into three-dimensional comoving coordinates.
The volume number density at the position of each point was defined as, 
$$ \Sigma_N = \frac{N}{4\pi d_N^3 /3},$$ 
where $d_N$ represents the comoving distance to the $N$th nearest neighbor. 
We used $N=5$, as the range $N=4-5$ is commonly employed (e.g., \citealt{2006MNRAS.373..469B}). 
Here, volume number densities were measured at 2 cMpc spacing.
The three-dimensional density contours were generated using {\sf plotly}\footnote{https://plotly.com} in Python.

\subsubsection{Group identification}
\label{sssec:method_fof}

Galaxy groups were identified using the Friends-of-Friends (FoF) algorithm, a method frequently utilized to search for groups in a distribution of galaxies or dark matter halos. This algorithm establishes a direct link between a particle and all other particles within a specified linking length $l$. These particles are then indirectly connected to the friends of the initially linked particles, forming a network of associations.
In this study, we employed linking lengths of $l_p = 1$ physical Mpc in the transverse direction and $l_z = 1000$ km s$^{-1}$ in the line-of-sight (redshift) direction. 
We identified FoF groups comprising $N\geq 10$ members. 

The halo mass of a group was estimated by applying the total stellar mass-to-halo mass ($M_{\rm 200}$) relation at $z\sim1$ found in GCLASS survey \citep{2014A&A...561A..79V}, where $\log (M_{\rm 200,\star}) = (12.44\pm0.04)+(0.59\pm0.10)\cdot[\log (M_{200})-14.5]$.
Here $M_{200}$ refers to the mass contained within $R_{200}$, the radius where the average density inside is 200 times the critical density.
The total stellar mass of a group was obtained by summing up the stellar masses of all group members within 1 Mpc in physical scale, twice $R_{200}$ of a halo with $M_{200}\sim 10^{14}~M_\odot$ in GCLASS survey at $z\sim1$ \citep{2014A&A...561A..79V}, from the group center.
We used a fixed radius because our FoF group members likely include galaxies in filaments connected to groups. 
We also estimated the halo masses using the total stellar masses of the group members within 2 Mpc.
A few objects were excluded due to contamination from QSOs.
This study focused exclusively on NBEs despite the potential presence of a non-negligible number of quiescent galaxies within a group. 
Consequently, the halo masses measured in this study are conservative lower limits.

The FoF groups were cross-matched with the {\it Chandra} COSMOS X-ray galaxy group catalog \citep{2019MNRAS.483.3545G}. 
This catalog contains extended X-ray sources likely associated with galaxies selected in the optical/NIR over our survey area. 
We selected X-ray sources that were detected with a significance of at least $5\sigma$ and located within 2 arcmin of the center of the FoF groups.
The halo masses $M_{\rm 200}$ listed in \citet{2019MNRAS.483.3545G} were estimated based on X-ray luminosity calculated with photometric redshifts. 
For the X-ray sources matched with groups, we recalculated the X-ray luminosity halo masses using our spectroscopic redshifts.

\subsubsection{Filaments extraction}
\label{sssec:method_filament}

Filamentary large-scale structures were identified using the Discrete Persistence Structure Extractor ({\sf DisPerSE}; \citealt{2011MNRAS.414..384S}).
This tool is extensively employed for filament identification, utilizing discrete Morse and persistence theories. Initially, the galaxy density field was constructed using {\sf delauney\_3D} within {\sf DisperSE}. The {\sf -btype smooth} option was implemented as a boundary control, as the volume of our target field is not a cube.
The {\sf DisperSE} utilizes Delaunay tessellation, wherein the density surrounding each vertex of the Delaunay complex is calculated through the Delaunay tessellation field estimator. Subsequently, the distribution of filaments was determined using the {\sf mse} command, which calculates Morse-Smale complexes and extracts filamentary structures. In this study, filaments were identified with a persistence threshold of $\sigma=5$.
The filaments were visualized using the {\sf skelconv} command. This process involves smoothing the network generated by {\sf mse} and converting it into a catalog of filaments. In this study, we applied smoothing with the {\sf -smooth 3} option in {\sf skelconv}.

\subsubsection{Voids extraction}
\label{sssec:method_voids}

In addition to the overdense regions, voids were searched in the survey volume. 
The past redshift surveys found voids that have a density contrast $\delta \rho/\rho$ lower than $-0.8\sim-0.9$, where $\rho$ is an average density and $\delta \rho/\rho=-1$ means a completely empty region (e.g., \citealt{2002ApJ...566..641H,2004ApJ...607..751H,2012MNRAS.421..926P}). 
Here, we selected voids satisfying $\delta \rho/\rho<-0.85$ in the volume density of O2Es.

The voids were searched avoiding 10 cMpc at the edge of the survey volume. 
To extract voids, volume number densities were measured in a similar manner as in section \ref{sssec:method_3d_nd} but with 4 cMpc spacing. 
A spacing larger than that used in section \ref{sssec:method_3d_nd} was applied to smooth the galaxy distribution.
First, we listed underdense points where the volume densities were less than 15\%~of the average density, corresponding to $\delta \rho/\rho<-0.85$ in the density of O2Es.
The neighboring underdense points were grouped by the FoF algorithm (section \ref{sssec:method_fof}), adopting a linking length $l=4$ cMpc.
We then extracted groups consisting of $\geq20$ underdense points, i.e. $\gtrsim 1000$ cMpc$^3$ in volume. 
The volume of a void was roughly calculated by multiplying the number of underdense points constituting a void by a $4\times4\times4$ cMpc$^3$ volume.

\begin{longtable}{lllllll}
  \caption{Summary of the FoF groups }\label{tab:groups}
\hline\noalign{\vskip3pt} 
Group ID & R.A.\footnotemark[$*$]  & Dec.\footnotemark[$*$] & $z_{\rm med}$\footnotemark[$*$] & $N_{\rm group}$\footnotemark[$\dag$] & $M_{\rm\star,<1Mpc}$($M_{200})$ & $M_{\rm\star,<2Mpc}$($M_{200}$)\\  [2pt] 
 & (deg) & (deg) &  &  & ($M_{\odot}$) & ($M_{\odot}$)\\  [2pt] 
\hline\noalign{\vskip3pt} 
\endfirsthead      
\hline\noalign{\vskip3pt} 
  Name & Value1 & Value2 & Value3 & Value4 & Value5 & Value6 \\  [2pt] 
\hline\noalign{\vskip3pt} 
\endhead
\hline\noalign{\vskip3pt} 
\endfoot
\hline\noalign{\vskip3pt} 
\multicolumn{2}{@{}l@{}}{\hbox to0pt{\parbox{160mm}{\footnotesize
\hangindent6pt\noindent
\hbox to6pt{\footnotemark[$*$]\hss}\unskip%
The median R.A., Declination, and redshift of a group. \\
\hbox to6pt{\footnotemark[$\dag$]\hss}\unskip%
The number of group members. \\
\hbox to6pt{\footnotemark[$\ddag$]\hss}\unskip%
The groups found at $>0.6$ deg from the survey area center or outside the redshift intervals $z=1.184-1.200$, $z=1.460-1.480$ and $z=1.592-1.612$.
}\hss}} 
\endlastfoot 
COS-NB816O2E-1           & 150.5420160 & 2.5278037 & 1.1973 & 41 & 1.8E+11 (3.0E+12) & 2.6E+11 (5.6E+12)\\
COS-NB816O2E-2           & 150.2258076 & 2.0030693 & 1.1904 & 27 & 9.7E+10 (1.1E+12) & 4.2E+11 (1.3E+13) \\
COS-NB816O2E-3\footnotemark[$\ddag$] & 149.8783187 & 2.2664826 & 1.1800 & 24 & 3.2E+10 (1.7E+11) & 1.1E+12 (1.4E+12) \\
COS-NB816O2E-4           & 150.6062406 & 2.3600686 & 1.1932 & 21 & 9.9E+10 (1.1E+12) & 1.1E+12 (1.3E+12)\\
COS-NB816O2E-5           & 150.3755895 & 2.0831590 & 1.1937 & 20 & 2.2E+11 (4.3E+12) & 2.5E+11 (5.3E+12) \\
COS-NB921O2E-1\footnotemark[$\ddag$] & 149.9788146 & 2.3336943 & 1.4587 & 95 & 3.5E+11 (9.5E+12) & 1.0E+12 (5.6E+13)\\
COS-NB921O2E-2\footnotemark[$\ddag$] & 149.4021876 & 2.2708385 & 1.4742 & 41 & 9.0E+10 (9.8E+11) & 4.0E+11 (1.2E+13)\\
COS-NB921O2E-3           & 149.8607878 & 2.1198243 & 1.4790 & 34 & 2.5E+11 (5.3E+12) & 3.3E+11 (9.0E+12) \\
COS-NB921O2E-4           & 149.5186018 & 2.1603114 & 1.4802 & 26 & 5.3E+11 (1.9E+13) & 6.8E+11 (3.0E+13) \\
COS-NB921O2E-5\footnotemark[$\ddag$] & 150.7929616 & 2.2003986 & 1.4696 & 25 & 3.0E+11 (7.4E+12) & 6.3E+11 (2.7E+13) \\
COS-NB921O2E-6           & 150.2975273 & 2.7460547 & 1.4652 & 25 & 1.8E+11 (3.1E+12) & 4.0E+11 (1.2E+13)\\
COS-NB921O2E-7\footnotemark[$\ddag$] & 150.3963860 & 1.5700392 & 1.4675 & 23 & 5.7E+10 (4.0E+11) & 1.7E+11 (3.0E+12)\\
COS-NB973O2E-1           & 150.2831947 & 2.2549435 & 1.6028 & 14 & 1.5E+11 (2.5E+12) & 3.5E+11 (9.3E+12)\\
COS-NB973O2E-2\footnotemark[$\ddag$] & 150.6966182 & 2.2045475 & 1.6122 & 12 & 2.2E+11 (4.4E+12) & 2.5E+11 (5.6E+12) \\
COS-NB973O2E-3           & 150.3021381 & 2.0152348 & 1.6034 & 10 & 2.5E+11 (5.5E+12) & 4.2E+11 (1.3E+13)\\
\end{longtable}

\begin{figure*}
 \begin{center}
  \includegraphics[width=\linewidth]{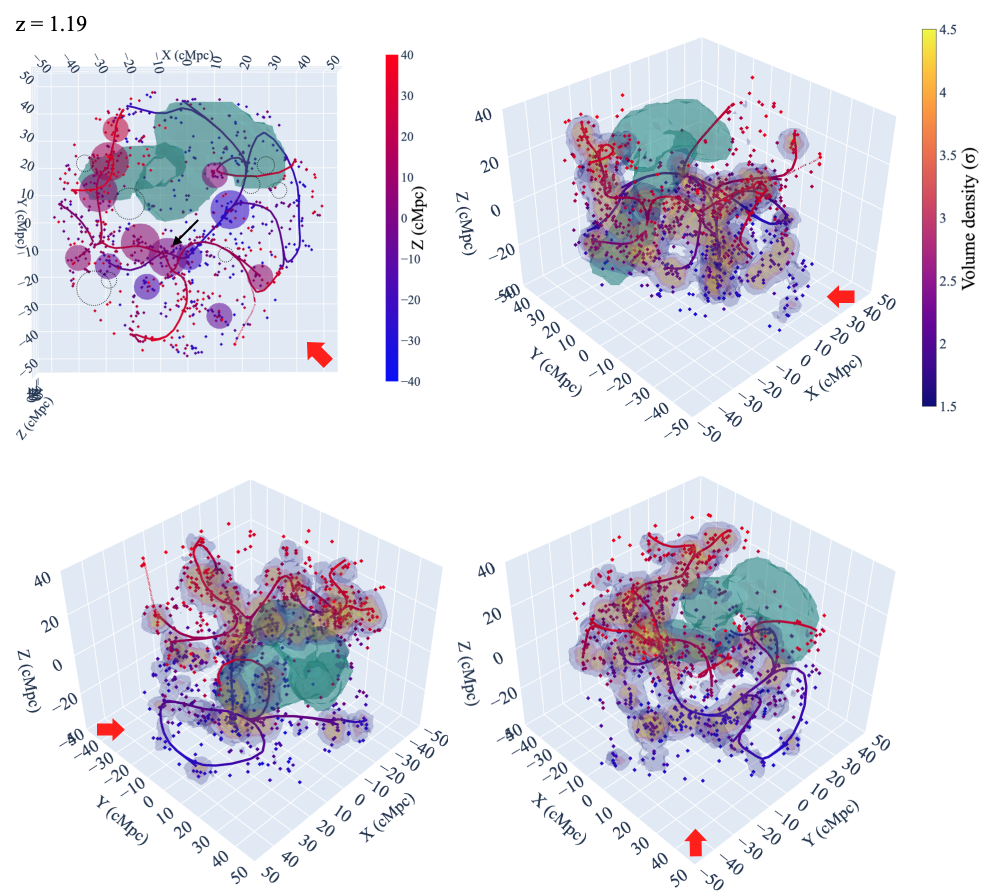} 
 \end{center}
\caption{The three-dimensional large-scale structure at $z=1.19$.
The {\it top left} panel shows the distributions of O2Es, groups, filaments, and voids at $z=1.19$, viewed from the line-of-sight direction. 
The other three panels show the distributions of O2Es, filaments, voids, and volume density contours.
The viewing angles of the four panels are different, but the red arrow in each panel points in the same direction. 
The points and solid curves show O2Es and filaments, respectively.
The dotted curves show filaments which are likely false detections, pointing to the boundary of the survey area, where no significant overdensity was observed. 
The small and large filled circles in the {\it top left} panel show groups consisting of $10-19$ members and $>20$ members, respectively.
The colors of O2Es, filaments, and groups represent the coordinate in the line-of-sight direction in a comoving scale, as shown in the color bar in the {\it top left} panel. 
The blue, magenta, and yellow colored contours show the 1.5, 3.0, and 4.5 $\sigma$ density contours corresponding to the color bar in the {\it top right} panel. 
The green-surfaced volumes show voids.
Both the volume density contours and void volumes were visualized using {\sf plotly} software.
The black arrow in the {\it top left} panel indicates the location of the richest group described in section \ref{ssec:fof}. 
The dotted open circles in the {\it top left} panel show the large bright star masks in the survey area. 
}\label{fig:disperse1}
\end{figure*}

\begin{figure*}
 \begin{center}
  \includegraphics[width=\linewidth]{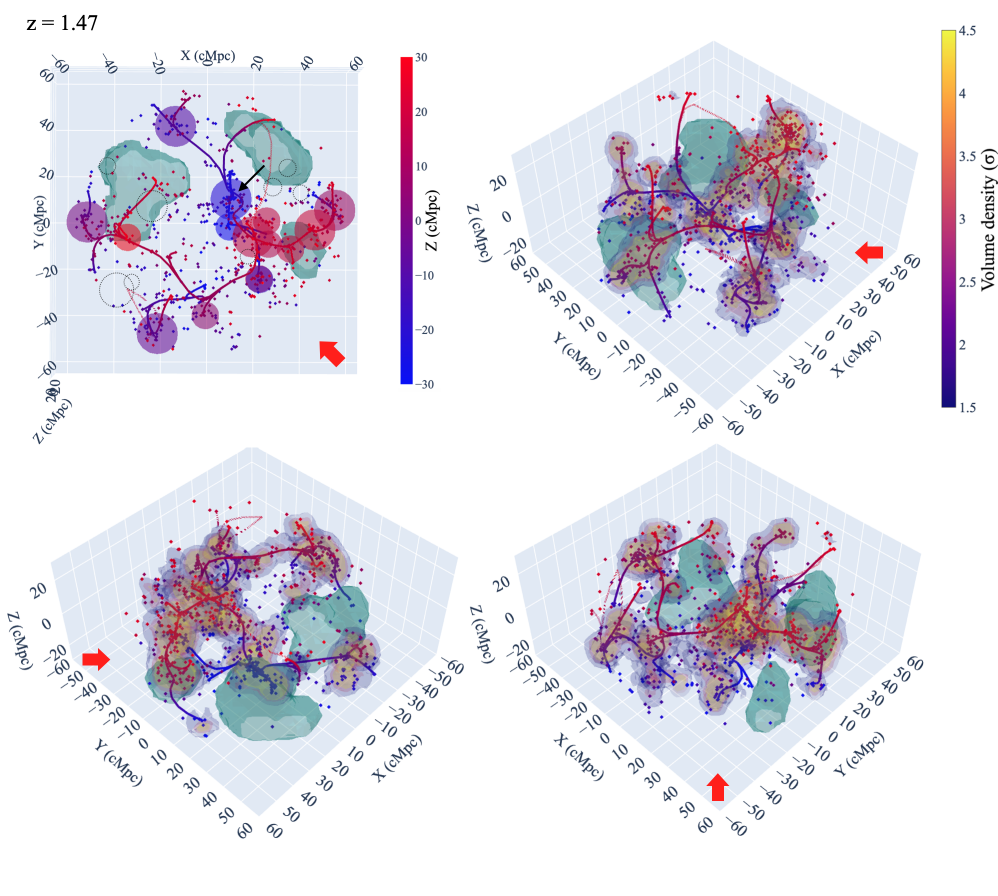} 
 \end{center}
\caption{ Similar to Figure \ref{fig:disperse1} but at $z=1.47$.  
}\label{fig:disperse2}
\end{figure*}

\begin{figure*}
 \begin{center}
  \includegraphics[width=\linewidth]{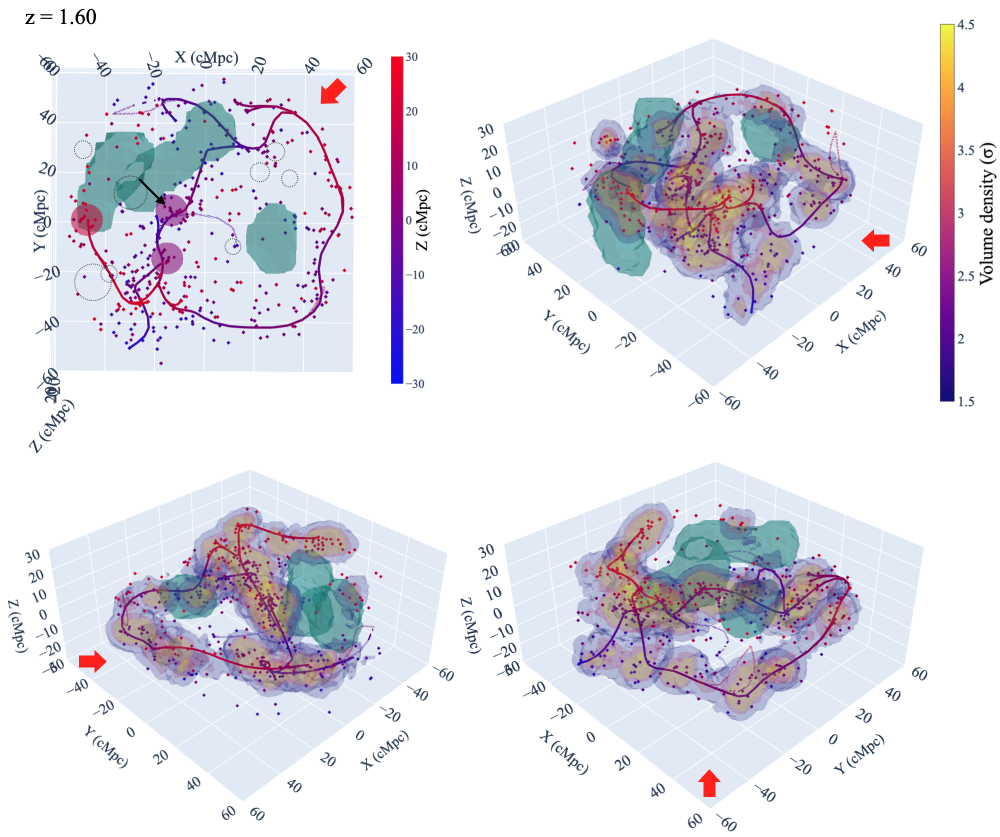} 
 \end{center}
\caption{ Similar to Figure \ref{fig:disperse1} but at $z=1.60$.  
}\label{fig:disperse3}
\end{figure*}

\begin{figure*}
 \begin{center}
  \includegraphics[width=\linewidth]{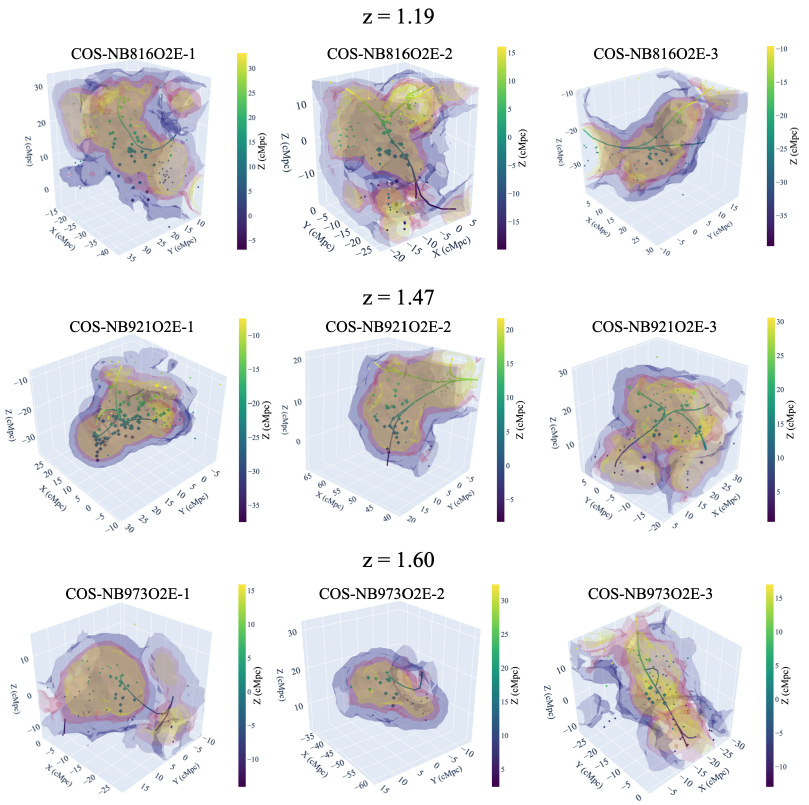} 
 \end{center}
\caption{ The distribution of O2Es within the three richest groups at each redshift in Table \ref{tab:groups} is depicted.
The large and small filled diamonds represent the group member and other O2Es.
The curves show filaments.
The color bars for O2Es and filaments correspond to the coordinate in the line-of-sight direction in a comoving scale.
The contours show the volume density in the color scale, the same as in Figure \ref{fig:disperse1}.
}\label{fig:groups1}
\end{figure*}

\subsection{FoF groups}
\label{ssec:fof}

Table \ref{tab:groups} lists the groups identified using the FoF algorithm.
The groups with a number of group members $N_{\rm group}\geq20$ are listed at $z=1.19$ and $z=1.47$, while groups with $N_{\rm group}\geq10$ are listed at $z=1.60$ because there were no groups with $N_{\rm group}\geq20$. 
The {\it top left} panels of Figures \ref{fig:disperse1}, \ref{fig:disperse2}, and \ref{fig:disperse3} shows the distributions of groups 
and Figure \ref{fig:groups1} presents the distributions of O2Es and filaments at the three richest groups at each redshift.
The black arrows in Figures \ref{fig:disperse1}, \ref{fig:disperse2}, and \ref{fig:disperse3} indicates the richest ($z=1.47$ and $z=1.60$) and/or X-ray detected group ($z=1.19$ and $z=1.47$) at each redshift.
The groups COS-NB816O2E-3, COS-NB921O2E-1, 2, 5, and 7, and COS-NB973O2E-2 are at $>0.6$ deg from the survey area center or outside the redshift intervals completely sampled. 
Thus, their richness and halo masses were likely underestimated.

The halo masses of the groups range from $1.2\times10^{12}$ to $5.6\times10^{13}~M_{\odot}$ if we consider group members within 2 physical Mpc of the group center.
They range from $0.2\times10^{12}$ to $1.9\times10^{13}~M_{\odot}$ if we use 1 physical Mpc as the clustercentric radius.
The halo masses measured by adopting 1 physical Mpc as a clustercentric radius are not likely reliable because rich groups with $\geq20$ members are unlikely to have halo masses below $10^{12}~M_\odot$, even though our sample is biased to low-mass galaxies.
As we discuss later, O2Es spread more widely than quiescent galaxies and favor outskirts of groups (e.g., \citealt{2014A&A...561A..79V,2014ApJ...796...65M}). 
Then a cluster mass cannot be well measured by counting O2Es within a small ($\sim1$ physical Mpc) clustercentric radius.

One group at each redshift, COS-NB816O2E-2, COS-NB921O2E-1, and COS-NB973O2E-1, has potential X-ray counterparts.
COS-NB816O2E-2 matches with an extended X-ray source, ID10064 in \citet{2019MNRAS.483.3545G}, 
which has an X-ray halo mass $M_{\rm 200} \sim4\times10^{13}~M_{\odot}$.
The halo mass based on total stellar mass within 2 Mpc of COS-NB816O2E-2 is of a similar order. 
Thus, this X-ray source plausibly associates with COS-NB816O2E-2.
Although COS-NB816O2E-1 is the richest group in O2Es, X-ray detected COS-NB816O2E-2 is plausibly the most massive group at $z=1.19$.
The richest group at $z=1.47$, COS-NB921O2E-1 matches with two extended X-ray sources, ID20161 and ID20228 in \citet{2019MNRAS.483.3545G}, 
which has an X-ray halo mass $M_{\rm 200c} \sim2\times10^{13}~M_{\odot}$ for each.
The sum of these halo masses is comparable to the halo mass based on the total stellar mass within 2 Mpc in this work.
Thus, COS-NB921O2E-1 can be associated with two extended X-ray sources.
COS-NB973O2E-1 matches with an extended X-ray source, ID10142 in \citep{2019MNRAS.483.3545G}, 
which has an X-ray halo mass $M_{\rm 200c} \sim4\times10^{14}~M_{\odot}$.
This value is too large compared to the halo mass based on the total stellar mass.
At this point, we cannot conclude that COS-NB973O2E-1 is an X-ray detected cluster. 

In summary, there is a lot of uncertainty in halo masses based on the total stellar masses of O2Es.
But if we adopt 2 Mpc as the clustercentric radius, a good order-of-magnitude estimate is obtained compared to the X-ray luminosity halo masses for the richest groups.
Plausibly, COS-NB816O2E-2 and COS-NB921O2E-1 represent the most massive cluster at each redshift interval in the survey area. 

\subsection{Filaments}
\label{ssec:filaments}

The distribution of the DisPerSE filaments is presented in Figures \ref{fig:disperse1}, \ref{fig:disperse2}, \ref{fig:disperse3}, and \ref{fig:groups1}.
The filaments extending towards the boundary of the survey area, where no significant overdensity was observed, are shown with dotted curves, as they are likely false detections.
The filaments extracted using the photometric catalog of O2Es \citep{2020PASJ...72...86H} are presented in Figure \ref{fig:disp_2d3d} in appendix.
The filaments extracted from the spectroscopic catalog are generally aligned with those identified in the photometric catalog.
On the other hand, the filaments extracted from the photometric catalog are often superpositions of filaments in the line-of-sight direction. 
Our study demonstrated that spectroscopic analysis is essential for the robust identification of filaments.

Most FoF groups are interconnected through filaments.
In particular, rich groups are connected to multiple filaments, as seen in Figure \ref{fig:groups1}.
COS-NB816O2E-2, the most massive group at $z=1.19$, is at the densest node in this redshift interval.
At $z=1.47$, all the rich groups are at the edge of the survey area and/or redshift coverage; filaments along these structures may be incompletely probed.
Besides incompleteness, COS-NB921O2E-1, the richest and X-ray-detected group at $z=1.47$, is presented as the node of the most significant filaments at this redshift interval.
At $z=1.60$, the connectivity of filaments to the groups was not clearly seen, maybe because of the low sample size to identify filaments robustly. 
In summary, we confirmed the strong connectivity of filaments for massive halos at $z=1.19$ and $1.47$, which has been demonstrated at lower redshifts in observations and simulations (e.g., \citealt{2019A&A...632A..49S,2019MNRAS.489.5695D,2020MNRAS.491.4294K,2021A&A...651A..56G,2026A&A...711A..35E}).

\subsubsection{Voids}
\label{ssec:voids}
 
The green-surfaced volumes in Figures \ref{fig:disperse1}, \ref{fig:disperse2}, and \ref{fig:disperse3} present the voids found in the survey volume.
Although we did not consider bright star masks in the analyses, 
a large fraction of the volumes of voids was certainly outside the area masked by bright stars. 

At $z=1.19$, one large void with a volume of $1.1\times10^4$ cMpc$^3$ was detected.
At $z=1.47$, voids with volumes of 1.5, 7.7, and 8.5 $\times10^3$ cMpc$^3$ were detected, and at $z=1.60$, voids with volumes of 1.9 and 9.7 $\times10^3$ cMpc$^3$ were detected. 
The sizes and volumes of these voids are smaller than voids found in the previous redshift surveys (e.g., \citealt{2002ApJ...566..641H,2004ApJ...607..751H,2012MNRAS.421..926P}); however, the voids in the PEGASUS certainly have a significantly low density contrast comparable to the voids found in previous studies.
As shown in the {\it top left} panels of Figures \ref{fig:disperse1}, \ref{fig:disperse2}, and \ref{fig:disperse3}, 
these voids were not always found as significant underdense regions on the two-dimensional maps of galaxies. 
Thus, our sample demonstrates the spectroscopic survey power of the PFS to identify voids which cannot be discovered by photometric surveys. 

\section{Discussion}

In this first paper of the PEGASUS project, we presented the survey performance and reproduction of large-scale structures with O2Es at $z>1.1$.
Many of the targets were successfully confirmed by detecting one or more emission-lines for each. 
One of the primary aims of the PEGASUS is to discuss the gas-phase metallicities of galaxies in diverse environments. 
As shown in Figure \ref{fig:2dimage}, several emission-line diagnostics are available for many of the targets. 
[Ne {\footnotesize III}]$\lambda3869$ is available for most targets in all redshift ranges. 
As our targets are low-mass galaxies with narrow line widths, [O {\footnotesize II}] doublets are generally well resolved at the spectral resolution of the PFS. 
Thus we can discuss the evolution of metallicities at $z\leq 1.6$ using the [Ne {\footnotesize III}]$\lambda3869$/[O {\footnotesize II}]$\lambda3727$ ratios \citep{2006A&A...459...85N} in upcoming papers. 
We showed that the flux losses from fibers have a large scatter by comparing the narrow-band and spectroscopic emission-line fluxes. 
Thus, to measure, e.g., SFR based on emission-lines, we need to correct the flux losses from fibers carefully. The narrow-band images taken with the HSC will help us correct for the fiber flux losses of our targets.

The spatial and velocity distributions of galaxies inform us how clusters and galaxies therein have evolved. 
Previously, GCLASS survey studied the spatial and velocity distributions of galaxies in massive clusters at $z\sim1$ \citep{2014A&A...561A..79V,2014ApJ...796...65M}.
The GCLASS clusters with $M_{200}=1-2\times10^{14}~M_\odot$ have velocity dispersions $\sigma_v=500-700$ km s$^{-1}$ and $R_{200}=0.5-1$ physical Mpc.
On the phase-space diagram in \citet{2014ApJ...796...65M}, cluster members with clustercentric radii smaller than $2 R_{200}$ are presented, and most of the member galaxies have clustercentric velocities lower than $2\sigma_v$.
Quiescent galaxies tend to be found at small clustercentric radii ($<0.5 R_{200}$) and have low clustercentric velocities at the core ($<1 \sigma_v$). 
Star-forming galaxies avoid the cluster core and are scattered over the extent of clusters.

\begin{figure*}
 \begin{center}
  \includegraphics[width=0.95\linewidth]{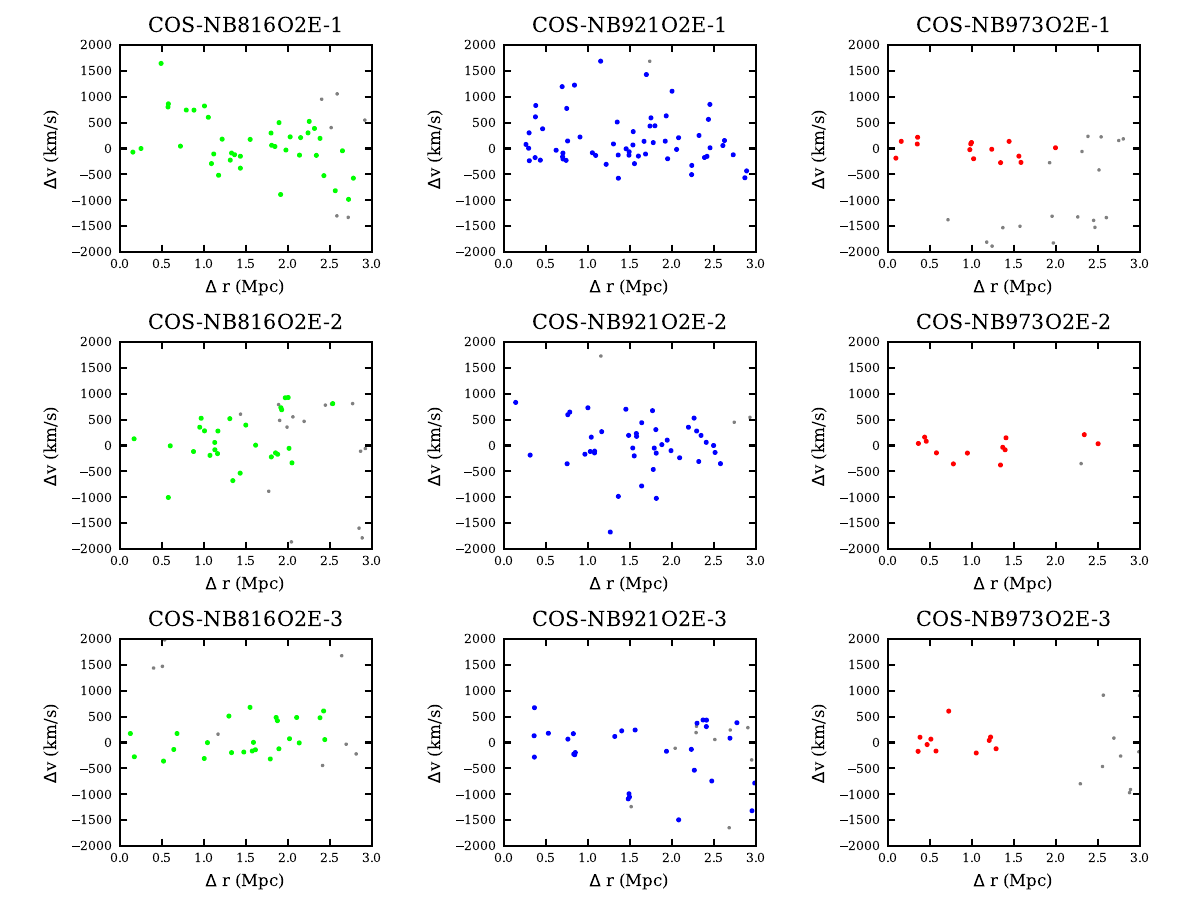} 
 \end{center}
\caption{The velocity vs. clustercentric radius phase space diagrams of the three richest groups at each redshift interval. In each panel, the velocities and spatial distributions are centered on the group centers in Table \ref{tab:groups}. The large colored circles represent O2Es within the groups, whereas the gray points indicate O2Es located outside the groups.
Note that COS-NB816O2E-3, COS-NB921O2E-1, and 2, and COS-NB973O2E-2 are sampled incompletely on one side of the redshift direction, as they are at the edge of the narrow-band survey coverage.
}\label{fig:groups2}
\end{figure*}

Figure \ref{fig:groups2} shows the velocity vs. clustercentric radius phase space diagrams for the richest groups found with the PEGASUS presented in Figure \ref{fig:groups1}. 
Note that COS-NB816O2E-3, COS-NB921O2E-1 and 2, and COS-NB973O2E-2 were sampled incompletely on one side of the redshift direction because they are at the edge of the narrow-band survey coverage.
The distribution of galaxies along the redshift axis reveals both their geometrical distances and peculiar velocities.
As we employed a linking length of $l_z = 1000$ km s$^{-1}$ in the redshift direction, corresponds to a physical distance of 1.5, 1.0, and 0.9 Mpc at $z=1.19$, $1.47$, and $1.60$, respectively, galaxies within a few 1000 km s$^{-1}$ of the group center can be selected as group members; however, most of O2Es at cluster centers have small velocity offsets $|\Delta v| <1000$ km s$^{-1}$. 
Thus, the extents of the groups along the line-of-sight direction are generally a few times smaller than
those in the transverse direction.
The velocity range of O2Es in COS-NB816O2E-2 and COS-NB921O2E-1, plausibly the most massive groups, is similar to that of star-forming galaxies in GCLASS clusters.
O2Es may trace star-forming galaxies that have recently fallen in and have not yet settled into the cluster core, as predicted in cosmological numerical simulations for O2Es \citep{2026MNRAS.551g1439O}.
From the above, O2Es do not probe the cores of clusters well; however, from a broader perspective, the rich groups of O2Es likely recover the locations of massive clusters.
We also need a structural analysis combined with quiescent galaxy population in forthcoming papers, as it was implemented photometrically in \citet{2026PASJ..tmp..197Y}. 

The connectivity of a halo with filaments has been studied by counting the number of filaments crossing a halo using photometric redshifts at $z\leq1.2$, and they reported three or fewer connections for halos with halo masses of $10^{13.5-14.0}~M_\odot$ \citep{2019A&A...632A..49S,2019MNRAS.489.5695D,2026A&A...711A..35E}.
As shown in Figure \ref{fig:groups1}, 
the richest groups at $z=1.19-1.47$ of the PEGASUS are connected to $4-5$ filaments, similar to the connectivity predicted in cosmological numerical simulations \citep{2019MNRAS.489.5695D,2020MNRAS.491.4294K}. 
Our pilot study demonstrated that a systematic spectroscopic survey for low-mass galaxies can robustly reproduce the distribution of filaments in simulations.

\section{Conclusion}

We conducted comprehensive spectroscopic observations of 9917 NBEs at $0.4\leq z\leq 1.6$ over a $\approx 1.8$ deg$^2$ area centered on the COSMOS field using the PFS on the Subaru Telescope. 
These NBEs, identified as HAEs, O2Es, and O3Es, were discovered by deep narrow-band imaging observations with the HSC and are representative of typical SFGs at these redshifts.
By utilizing emission-line fluxes derived from narrow-band excesses, this catalog allows us to perform effective spectroscopic observations for low-mass galaxies with high completeness at several redshift intervals. 
Between 86\% and 99\% of these targets were successfully confirmed by detecting one or more emission-lines for each.
Multiple emission-lines were identified in the majority of the targets. 
However, the detection rate of [N {\footnotesize II}]$\lambda\lambda6550,6585$ remains low, attributable to the performance of the NIR-arm and maybe the low-metallicity characteristics of the targets.
At $z=1.19-1.60$, the three-dimensional large-scale structures were examined. 
All galaxy groups are located at the nodes of filamentary structures, with the most massive groups situated at the densest intersections of these filaments. 
Voids that were not initially identified in the photometric analyses were extracted from the three-dimensional density maps.
This observation effectively corroborates the cosmic web as anticipated by cosmological numerical simulations.
Note that the targets of this work are biased to emission-line galaxies and the survey area is limited. The unbiased wide-field survey of the PFS-SSP and our PEGASUS project will complement each other. 
In our forthcoming publications, we will integrate our spectroscopic data with extensive multi-wavelength observations in the COSMOS field to investigate the evolution of galaxies in diverse environments in detail.

\begin{ack}

This research relies fundamentally on the narrow-band emission-line galaxy catalog generated by \citet{2020PASJ...72...86H}.
We acknowledge their great efforts in the construction of this catalog.

The instrument `\={O}nohi`ula - Prime Focus Spectrograph (PFS) including both hardware
and software was developed by the PFS collaboration consisting of over 25 institutes
across seven countries (in alphabetical order, Brazil, China, France, Germany, Japan,
Taiwan, and the United States), where the technical activities were conducted by (in
alphabetical order) Academia Sinica Institute of Astronomy and Astrophysics (Taiwan),
California Institute of Technology, Johns Hopkins University, Kavli Institute for the
Physics and Mathematics of the Universe in the University of Tokyo (Kavli IPMU),
Laboratoire d`Astrophysique de Marseille, Laboratório Nacional de Astrofísica (Brazil),
Max-Planck-Institut für Astrophysik, Max-Planck-Institut für extraterrestrische Physik,
NASA Jet Propulsion Laboratory, National Astronomical Observatory of Japan (NAOJ),
Princeton University, and and Universidade de São Paulo under the oversight by Project
Office hosted by Kavli IPMU (later NAOJ). There were also essential commitments from
academic and industrial partners such as Durham University (United Kingdom) and and
Bertin Technologies (France).
This work is based (in part) on data collected at the Subaru Telescope, which is
operated by the National Astronomical Observatory of Japan. We are honored and
grateful for the opportunity of observing the Universe from Maunakea, which has the
cultural, historical, and natural significance in Hawaii.
We appreciate the development and operation of PFS Science Platform by
Subaru Telescope and Astronomy Data Center at NAOJ which enables access to
both PFS and HSC data and various analyses on the server side.

The HSC collaboration includes the astronomical communities of Japan and
Taiwan, and Princeton University. The HSC instrumentation and software
were developed by the National Astronomical Observatory of Japan (NAOJ),
the Kavli Institute for the Physics and Mathematics of the Universe (Kavli
IPMU), the University of Tokyo, the High Energy Accelerator Research
Organization (KEK), the Academia Sinica Institute for Astronomy and
Astrophysics in Taiwan (ASIAA), and Princeton University. Funding was
contributed by the FIRST program from the Japanese Cabinet Office, the
Ministry of Education, Culture, Sports, Science and Technology (MEXT), the
Japan Society for the Promotion of Science (JSPS), Japan Science and
Technology Agency (JST), the Toray Science Foundation, NAOJ, Kavli IPMU,
KEK, ASIAA, and Princeton University.

This work has made use of data from the European Space Agency (ESA)
mission Gaia (https://www.cosmos.esa.int/gaia), processed by the Gaia Data
Processing and Analysis Consortium (DPAC,
https://www.cosmos.esa.int/web/gaia/dpac/consortium). Funding for the
DPAC has been provided by national institutions, in particular the institutions
participating in the Gaia Multilateral Agreement.

The Pan-STARRS1 Surveys (PS1) and the PS1 public science archive have
been made possible through contributions by the Institute for Astronomy, the
University of Hawaii, the Pan-STARRS Project Office, the Max Planck
Society and its participating institutes, the Max Planck Institute for
Astronomy, Heidelberg, and the Max Planck Institute for Extraterrestrial
Physics, Garching, The Johns Hopkins University, Durham University, the
University of Edinburgh, the Queen’s University Belfast, the
Harvard-Smithsonian Center for Astrophysics, the Las Cumbres Observatory
Global Telescope Network Incorporated, the National Central University of
Taiwan, the Space Telescope Science Institute, the National Aeronautics and
Space Administration under grant No. NNX08AR22G issued through the
Planetary Science Division of the NASA Science Mission Directorate, the
National Science Foundation grant No. AST-1238877, the University of
Maryland, Eotvos Lorand University (ELTE), the Los Alamos National
Laboratory, and the Gordon and Betty Moore Foundation.

\end{ack}

\section*{Funding}
 This work was supported by JSPS KAKENHI grant No. 22K21349 (TK, YK), 23K25911 (TN), 24H00002 (MK, TK), 24K17084 (YL), 25H00663 (MK, YL),  25K01032 (MK, RS), 25K07361 (MO), 26H02070 (MK, YK), and 26K17200 (YS).
 This work was also supported by JSPS Core-to-Core Program (grant number: JPJSCCA20210003), SUPER-IRNET.

\section*{Data availability} 
 The target catalog of this study is available at https://hsc-release.mtk.nao.ac.jp/doc/index.php/sample-page/pdr2/.
 The raw and reduced data underlying this study will be available at the data archive system of the National Astronomical Observatory of Japan. 
 We will publish a spectroscopic catalog of the PEGASUS in the forthcoming papers. 

\appendix 
\section{Requested exposure times}

Table \ref{tab:description1} summarizes the exposure time requested for each target category. 
The requested exposure times for NBEs depend on the wavelength of the primary emission-lines, i.e., the narrow-band filters used for their extraction, and emission-line fluxes measured with the narrow-band filters in \citet{2020PASJ...72...86H}.

\begin{longtable}{lcccc}
  \caption{Requested exposure time}\label{tab:description1}
\hline\noalign{\vskip3pt} 
& Exptime (1)\footnotemark[$*$] & Exptime (2)\footnotemark[$*$] & Exptime (3)\footnotemark[$*$] & Exptime (4)\footnotemark[$*$]\\ [2pt] 
 & (hr) & (hr) & (hr) & (hr) \\ [2pt] 
\hline\noalign{\vskip3pt} 
\endfirsthead      
\hline\noalign{\vskip3pt} 
  Name & Value1 & Value2 & Value3 & Value4   \\  [2pt] 
\hline\noalign{\vskip3pt} 
\endhead
\hline\noalign{\vskip3pt} 
\endfoot
\hline\noalign{\vskip3pt} 
\multicolumn{2}{@{}l@{}}{\hbox to0pt{\parbox{100mm}{\footnotesize
\noindent
\hbox to6pt{\footnotemark[$*$]\hss}\unskip%
Exptime (1), (2), (3), and (4) describe the exposure times for the targets with $F_{\rm NB} > 10,~6-10,~3-6,\&~2-3\times10^{-17}$ erg s$^{-1}$ cm$^{-2}$, respectively, for {\it NB527} to {\it NB973} NBEs.
The exposure times for all {\it NB1010} NBEs are described as Exptime (1).
}\hss}} 
\endlastfoot 
{\it NB527}  & 0.25 & 0.5 & 1.75 & 3.0 \\
{\it NB718}  & 0.5 & 1.0 & 3.0 & 4.0 \\
{\it NB816}  & 0.5 & 1.0 & 3.0 & 4.0 \\
{\it NB921}  & 0.5 & 1.0 & 3.0 & 4.0 \\
{\it NB973}  & 0.25 & 0.5 & 1.25 & 3.0 \\
{\it NB1010} & 2.0  & ... & ... & ...\\
\end{longtable}

\section{The redshift distribution}

\begin{figure*}
 \begin{center}
  \includegraphics[width=0.95\linewidth]{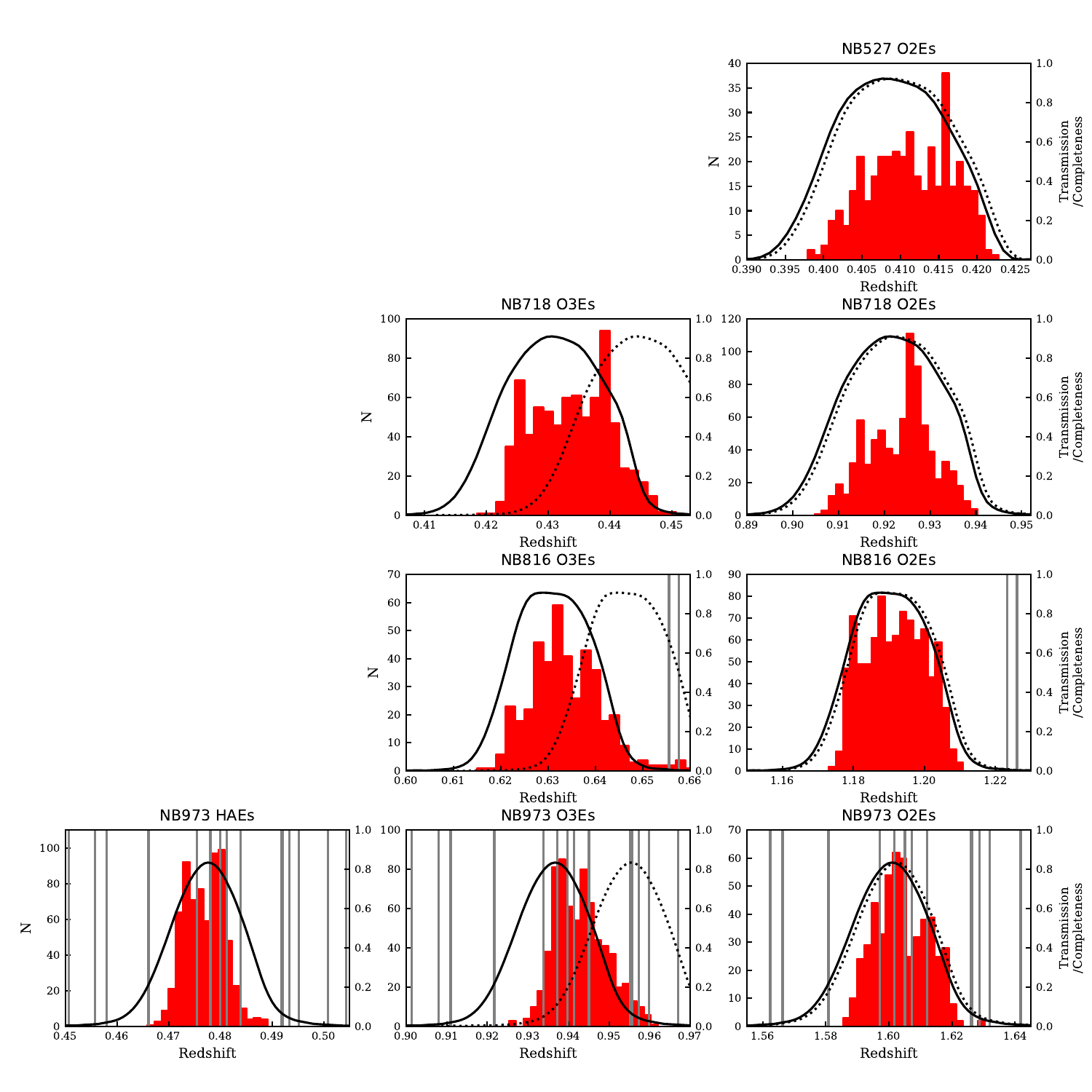} 
 \end{center}
\caption{Similar to Figure \ref{fig:zdist1} but the rows are NB527, NB718, NB816, and NB973 NBEs from top to bottom. 
}\label{fig:zdistmerged}
\end{figure*}

Figures \ref{fig:zdistmerged} present the redshift distributions of NBEs in the same manner as Figure \ref{fig:zdist1}.

\section{Emission-line fluxes measured with narrow-band and PFS}

\begin{figure*}
 \begin{center}
  \includegraphics[width=0.95\linewidth]{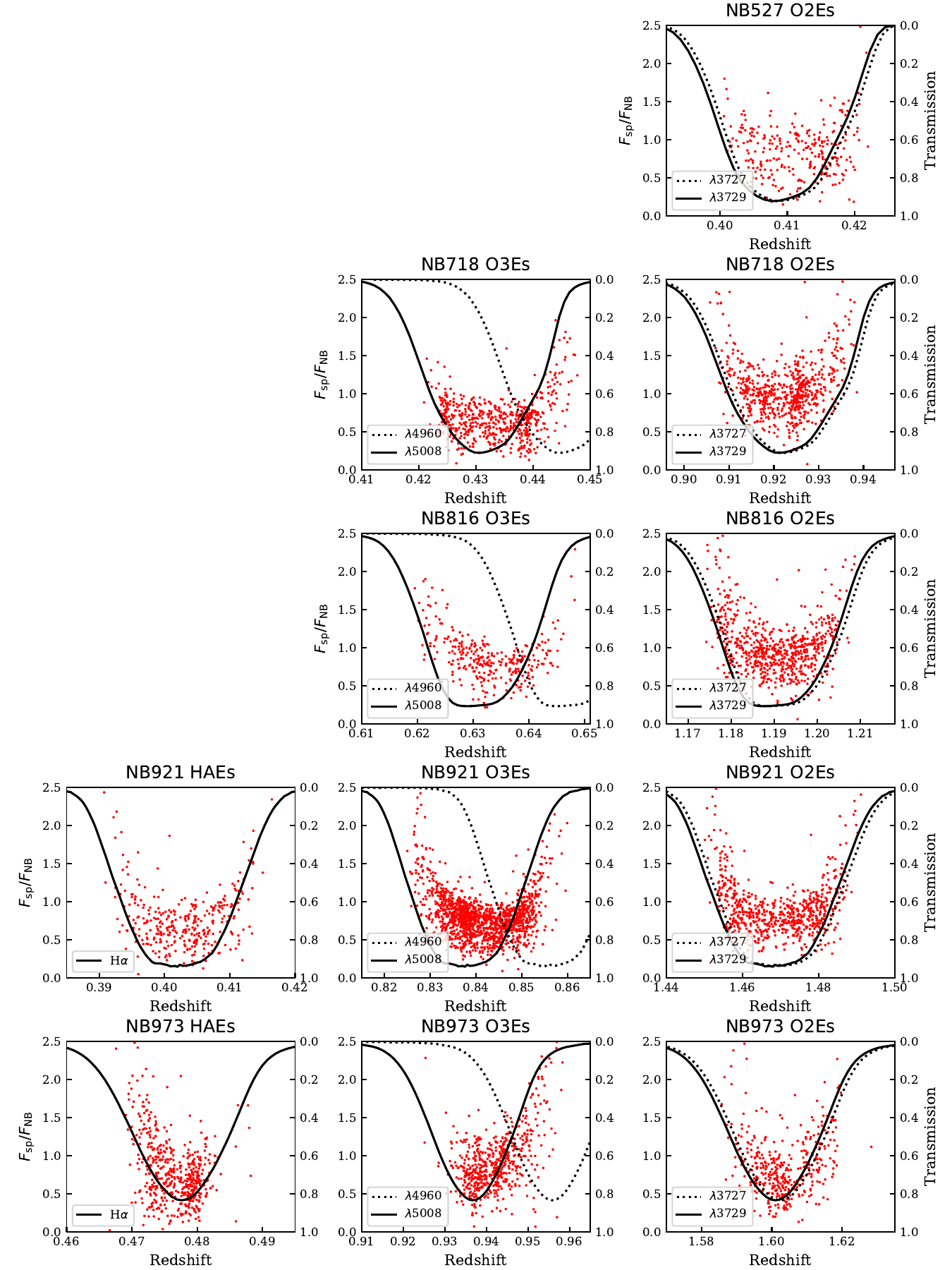} 
 \end{center}
\caption{The $Fl_{\rm sp}$/$Fl_{\rm NB}$ of NBEs. 
The panels are {\it NB527}, {\it NB718}, {\it NB816}, {\it NB921}, and {\it NB973} from top to bottom, and HAEs, O3Es, and O2Es from left to right.
The red points indicate spectroscopically confirmed NBEs.
For the HAEs, the black solid curves show the transmission curve of a filter scaled for H$\alpha$.
For O3Es, the black dotted and solid curves show the transmission curves of a filter scaled for [O {\footnotesize III}] $\lambda4960$ and $\lambda5008$, respectively.
For O2Es, the black dotted and solid curves show the transmission curves of a filter scaled for [O {\footnotesize II}] $\lambda3727$ and $\lambda3730$, respectively.
}\label{fig:fluxloss}
\end{figure*}

Figure \ref{fig:fluxloss} shows the $Fl_{\rm sp}$/$Fl_{\rm NB}$ of the {\it NB527} O2Es, {\it NB718} O3Es and O2Es, {\it NB816} O3Es and O2Es, {\it NB921} HAEs, O3Es and O2Es, and {\it NB973} HAEs, O3Es and O2Es.

\section{Filament extraction on the photometric catalog of O2Es}
\label{sec:disp_2d3d}

\begin{figure*}
 \begin{center}
  \includegraphics[width=0.95\linewidth]{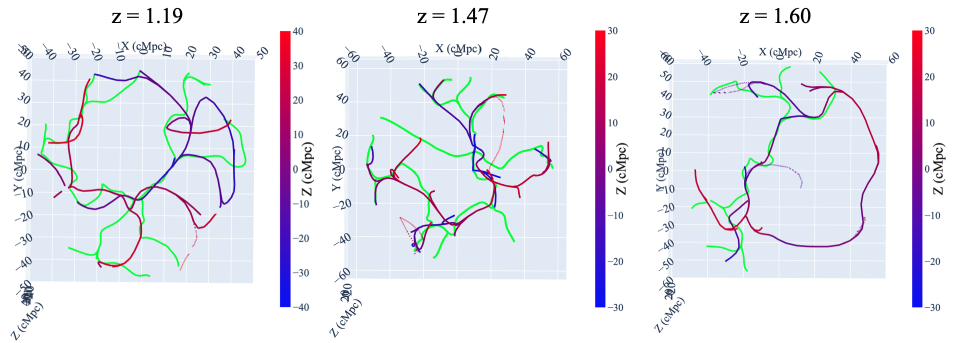} 
 \end{center}
\caption{The comparison of the filaments extracted from the photometric and spectroscopic catalogs at $z=1.19$, $1.47$, and $1.60$, from left to right. 
The blue-to-red colored curves are the same as the filaments based on the spectroscopic catalog in Figures \ref{fig:disperse1}, \ref{fig:disperse2}, and \ref{fig:disperse3}. 
The thick green curves show the filaments which were extracted from the photometric catalog of O2Es \citep{2020PASJ...72...86H} at each redshift interval.}\label{fig:disp_2d3d}
\end{figure*}

We compared the filaments extracted from the photometric catalog of O2Es and our spectroscopic catalog of O2Es at $z=1.19$, $1.47$, and $1.60$ in Figures \ref{fig:disp_2d3d}.
We used the R.A. and Dec. of O2Es at $z=1.19$, $1.47$, and $1.60$ in \citet{2020PASJ...72...86H} in the target list of the PEGASUS. 
The extraction of filaments from the photometric catalog of O2Es was performed in the same manner as that performed with the spectroscopic catalog in section \ref{sssec:method_filament}, but the galaxy density fields were constructed based on the two-dimensional coordinates using {\sf delauney\_2D} of {\sf DisPerSE}.

\end{document}